\newif\ifsubmode
\submodefalse

\newif\ifprintfig
\printfigtrue

\newif\ifemulate
\emulatetrue

\ifsubmode
  \documentclass[12pt,preprint]{aastex}
  \received{}
  \accepted{}
  \journalid{}{}
  \articleid{}{}
\else
   \documentclass{emulateapj}
   \submitted{{\it Accepted to the Astronomical Journal on July 15, 2011}}
\fi

\usepackage{longtable}

\newcommand{\kms}{\,km~s$^{-1}$}

\def\lesssim{\mathrel{\hbox{\rlap{\hbox{\lower4pt\hbox{$\sim$}}}\hbox{$<$}}}}
\def\gtrsim{\mathrel{\hbox{\rlap{\hbox{\lower4pt\hbox{$\sim$}}}\hbox{$>$}}}}

\def\spose#1{\hbox to 0pt{#1\hss}}
\def\simlt{\mathrel{\spose{\lower 3pt\hbox{$\mathchar"218$}}
     \raise 2.0pt\hbox{$\mathchar"13C$}}}
\def\simgt{\mathrel{\spose{\lower 3pt\hbox{$\mathchar"218$}}
     \raise 2.0pt\hbox{$\mathchar"13E$}}}

\shorttitle{Properties of Segue 3}
\shortauthors{Fadely~et~al.}

\begin{document}

\title{Segue 3: An Old, Extremely Low luminosity Star Cluster in the Milky Way's Halo}

  \author{Ross Fadely\altaffilmark{1}, Beth Willman\altaffilmark{1},
 Marla Geha\altaffilmark{2}, Shane Walsh\altaffilmark{3},\\ Ricardo R. Mu\~noz\altaffilmark{2,4},
 Helmut Jerjen\altaffilmark{5}, Luis C. Vargas\altaffilmark{2}, and Gary S. Da Costa\altaffilmark{5}}

\altaffiltext{1}{Department of Astronomy, Haverford
  College, Haverford, PA 19041; rfadely@haverford.edu, bwillman@haverford.edu}
\altaffiltext{2}{Astronomy Department, Yale University, New Haven, CT
  06520}
\altaffiltext{3}{Magellan Fellow; Australian Astronomical Observatory/Las Campanas Observatory}
\altaffiltext{4}{Departamento de Astronom\'ia, Universidad de Chile, Casilla 36-D, Santiago, Chile}
\altaffiltext{5}{Research School of Astronomy \& Astrophysics, Australian National University}

\ifsubmode\else
  \ifemulate\else
     \clearpage
  \fi
\fi


\ifsubmode\else
  \ifemulate\else
     \baselineskip=14pt
  \fi
\fi

%
%
\begin{abstract}
\renewcommand{\thefootnote}{\fnsymbol{footnote}}

We investigate the kinematic and photometric properties of the Segue 3 Milky Way companion using Keck/DEIMOS 
spectroscopy and Magellan/IMACS $g$ and $r$-band imaging.  Using maximum likelihood methods to analyze the 
photometry, we study the structure and stellar population of Segue 3.  We find the half--light radius of Segue 3 is 
$26''\pm5''$ ($2.1\pm0.4$ pc, for a distance of 17 kpc) and the absolute magnitude is a mere $M_V=0.0\pm0.8$ mag, 
making Segue 3 the least luminous old stellar system known.  We find Segue 3 to be consistent with a single stellar 
population, with an age of $12.0^{+1.5}_{-0.4}$ Gyr and an [Fe/H] of $-1.7^{+0.07}_{-0.27}$.  Line--of--sight velocities 
from the spectra are combined with the photometry to determine a sample of 32 stars which are likely associated with 
Segue 3.  The member stars within three half--light radii have a velocity dispersion of $1.2\pm2.6$ \kms.  Photometry 
of the members indicates the stellar population has a spread in [Fe/H] of $\lesssim0.3$ dex.  These facts, together 
with the small physical size of Segue 3, imply the object is likely an old, faint stellar cluster which contains no 
significant dark matter.  We find tentative evidence for stellar mass loss in Segue 3 through the eleven candidate 
member stars outside of three half-light radii, as expected from dynamical arguments.  Interpretation of the data 
outside of three half--light radii, is complicated by the object's spatial coincidence with a previously known halo 
substructure, which may enhance contamination of our member sample.
				   
\end{abstract}


%
%
\section{Introduction}
\label{sec:intro}

Segue 3 (hereafter Seg\,3) is a recently discovered Milky Way satellite, initially estimated to have an extremely low 
luminosity ($M_{\rm V} \sim -1.2$, $\sim$ 250 $L_{\Sun}$), a half-light radius of $\sim$ 3 pc, and a distance of 
$\sim$17 kpc \citep{belokurov10a}.  Five other Milky Way satellites are known to have comparably small luminosities
and scale sizes: Whiting 1 ($M_{\rm V} = -2.46$), Pal 1 ($M_{\rm V} = -2.52$), AM 4 ($M_{\rm V} \sim -1.8$), Koposov 
1 ($M_{\rm V} = -1.35$), and Koposov 2 ($M_{\rm V} \sim -0.35$) \citep[][2010 edition]{harrisGCcat}. All six of these 
objects are relative outliers in size and luminosity from other known Milky Way satellites, with luminosities a factor of 
three lower than those of the other 112 Milky Way satellites with half-light radii smaller than 5 pc. 

These six extreme satellites lie at distances ranging from 17 kpc (Segue 3, Pal 1) to 40 kpc or more (Koposov 1 and 
2).  Despite these halo distances, one explanation for their anomalously low luminosities is stellar mass loss owing 
to dynamical evolution, such as tidal stripping or tidal shocking.  At their current stellar masses and sizes, the 
evaporation timescales of these clusters are also shorter than the observed ages of the systems (see e.g. discussion 
in \citealt{koposov07b}), suggesting that evaporation is playing a major role in their evolution.  The hypothesis that 
these satellites are experiencing substantial stellar mass loss is supported by observations: Strong evidence for 
extra-tidal stars has been found in Pal 1, AM 4, and Whiting 1 \citep{ carraro09a, carraro07a,NO10a}. (Koposov 1 
and 2 haven't yet been studied thoroughly enough to confirm or rule out the presence of extra-tidal stars.)  Most of 
these six ultra-low luminosity satellites have been (at least tentatively) associated with larger scale stellar streams in 
the halo. For example, Whiting 1 and Koposov 1 have been associated with the Sgr stream \citep{carraro07a,
koposov07b}, AM 4 possibly with the Sgr stream \citep{carraro09a}, Pal 1 possibly with the Galactic Anticenter Stellar 
Structure \citep[GASS; ][]{frinchaboy04} or Canis Major \citep{forbes10a}, and Segue 3 possibly with the 
Hercules-Aquila cloud \citep{belokurov10a}.

Perhaps all six of these $M_V > -2.5, r_{1/2} < 5$ pc satellites are stripped down versions of more luminous objects, 
and can provide insight into the evolution of the Milky Way's satellite population and into the build-up of the halo. 
Because of their small physical sizes and luminosities alone, they have all been classified as globular clusters in the 
literature.  Their central surface brightnesses are higher than those of the known Milky Way dwarfs, so they are not 
simply stripped versions of those objects.  Although this circumstantial evidence supports a model where these 
outliers may instead be stripped versions of more luminous globular clusters, no robust classification or detailed 
kinematic study of these objects has yet been done.

We aim to improve our understanding of these unusual Milky Way satellites by conducting a detailed kinematic study 
of Segue 3, the first to be done for any comparable object. In particular, we study Magellan/IMACS photometry in $g$ 
and $r$ of Segue 3 to investigate its structural parameters and star formation history.  We combine this photometry 
with Keck/DEIMOS spectroscopy to obtain a robust member sample.  We use these spectroscopic members to study 
Seg 3's internal kinematics and to look for evidence of the extra-tidal stars seen in similar clusters. We use its internal
kinematics, improved structural parameters, and star formation history to infer a star cluster classification for Seg\,3.

%
%
\section{Data}
\label{sec:data}

\subsection{Photometry}
\label{ssec:phot}

We observed Segue 3 using Magellan/IMACS during engineering time on 2010 August 21 with bright conditions 
and excellent $\sim0.3''$ seeing. We used the IMACS in f/4 mode, giving a pixel scale of $0.111''/$pixel and a
$15.5'\times15.5'$ field of view.  We obtained a total integration time of 1620s in each of the $g$ and $r$ Sloan filters, 
using observations which were dithered across the sky.

Individual chip images were reduced and stacked to create a weighted mosaic using {\tt SExtractor}, {\tt SCAMP} and 
{\tt SWarp} \citep{bertin96a,bertin02a,bertin06a}, with SDSS DR7 providing the astrometric calibration. Normalized 
flats were used to weight the images, with zero weightings given for known defects. The stacked mosaics were then 
photometered using the stand--alone {\tt DAOPHOT II}/{\tt Allstar} package \citep{stetson87}.  

The photometry was calibrated to SDSS by astrometrically matching the stars and then fitting for the difference 
between instrumental and SDSS magnitudes and colors.  To robustly determine the calibration parameters, we used 
a 1000 iteration bootstrap method which calibrated our detections to SDSS sources within our region of interest in the 
color--magnitude diagram: sources with $g$ and $r$ magnitudes between 18.0 and 22.5, and $0.1<g-r<0.5$. For 
each iteration, we first use an iterative $3\sigma$ clip to remove outliers and then do an uncertainty weighted fit for the 
zeropoints and color terms. We adopt the median values of the bootstrap analysis for the calibration, along with their 
standard deviations. The calibration we determine is:

\begin{eqnarray}
g &=& g_{\rm inst} + 1.588 \pm0.017 - 0.079\pm0.016(g_{\rm inst}-r_{\rm inst}), \nonumber \\ 
r &=& r_{\rm inst} + 1.887 \pm0.014 - 0.018\pm0.017(g_{\rm inst}-r_{\rm inst}). \nonumber
\end{eqnarray}

After calibration, we conducted Monte Carlo tests to assess the completeness of our photometric data.  Specifically, 
completeness limits were calculated by injecting the mosaic image with a $250\times250$ grid of artificial stars, 
spaced 30 pixels apart in $x$ and $y$ directions plus a small random offset.  We repeated this injection 24 times for a 
total of 1,500,000 artificial stars in each filter. For each star, we randomly assign a magnitude between 18 and 25 and 
subsequently detected the stars using the same criteria as the real data.  We set valid detections as those with a 
position within 1 pixel of the input position, and measure the fraction recovered as a function of magnitude. We find the 
90\% completion limits for our $g$ and $r$-band data are 23.7 and 23.9, respectively.

\subsection{Spectroscopy}
\label{ssec:spect}

The spectroscopic data were taken with the Keck~II 10-m telescope and the DEIMOS spectrograph \citep{faber03a}.  
To select spectroscopic targets, we used the SDSS DR7 photometry, because the IMACS photometry discussed in 
Section \ref{ssec:phot} was not available at the time.  Candidate Segue\,3 member stars were identified by 
comparison to a Padua theoretical isochrone for an age of 13\,Gyr and a metallicity of [Fe/H]$= -2.3$ \citep{marigo08}.  
Stars within 0.1\,mag of this isochrone and brighter than $r = 22$ were given the highest priority for spectroscopy; stars 
within 0.2\,mag of the isochrone were given lower priority, all other stars observed were selected in order to fill in the 
mask.  Slitmasks were created using the DEIMOS {\tt dsimulator} slitmask design software.

Two Keck/DEIMOS multislit masks were observed on November 16, 2009 and a third mask on May 16, 2010. The 
masks were observed for one hour each with the 1200~line~mm$^{-1}$\,grating covering a wavelength region 
$6400-9100\mbox{\AA}$.  The spectral dispersion of this setup is $0.33\mbox{\AA}$, and the resulting spectral 
resolution, taking into account the anamorphic distortion, is $1.37\mbox{\AA}$ (FWHM, equivalent to 47\kms\ at the 
Ca II triplet). The spatial scale is $0.12''$~per pixel and slitlets were $0.7''$ wide.  The minimum slit length was $4''$ 
which allows adequate sky subtraction; the minimum spatial separation between slit ends was $0.4''$ (three pixels).

Spectra were reduced using a modified version of the {\tt spec2d} software pipeline (version~1.1.4) developed by the 
DEEP2 team at the University of California--Berkeley for that survey. A detailed description of the reductions can be 
found in \citet{simon07a}.  The final one-dimensional spectra are rebinned into logarithmic wavelength bins with 
15\kms\ per pixel.  Radial velocities were measured by cross-correlating the observed science spectra with a series 
of high signal--to--noise stellar templates.  We calculate and apply a telluric correction to each science spectrum by 
cross--correlating a hot stellar template with the night sky absorption lines following the method in \citet{sohn07}.  
The telluric correction accounts for the velocity error due to mis--centering the star within the $0.7''$ slit caused by
small mask rotations or astrometric errors.  We apply both a telluric and heliocentric correction to all velocities 
presented in this paper.

We determine the random component of our velocity errors using a Monte Carlo bootstrap method.  Noise is added to 
each pixel in the one--dimensional science spectrum, we then recalculate the velocity and telluric correction for 1000 
noise realizations.  Error bars are defined as the square root of the variance in the recovered mean velocity in the 
Monte Carlo simulations.  The systematic contribution to the velocity error was determined by \citet{simon07a} to be
2.2\kms\ based on repeated independent measurements of individual stars.  The systematic error contribution is 
expected to be constant as the spectrograph setup and velocity cross--correlation routines are identical.  We add the 
random and systematic errors in quadrature to arrive at the final velocity error for each science measurement.  Radial 
velocities were successfully measured for 163 of the 205 extracted spectra across.  The fitted velocities were visually
inspected to ensure reliability.  Our final sample consists of 149 measurements of 132 unique stars, presented in 
Tables \ref{tab:mem} and \ref{tab:nonmem}.  Of the 17 stars with repeat measurements, we identify one binary star 
candidate (ID 9, Table \ref{tab:mem}), with velocities which differ by more than $3\sigma$ between observation epochs.  
The remaining stars exhibit velocities consistent within $2\sigma$.  For our analysis below, we use a combined velocity 
for all stars with repeat measurements, but exclude the identified binary when calculating the velocity dispersion of Seg\,3.

%
%
\section{A Spectroscopic Sample of Segue 3 Stars}
\label{sec:members}

\begin{figure*}
\centering
\includegraphics[clip=true, trim=1.2cm 12.1cm 2.0cm 2.cm,width=12cm]{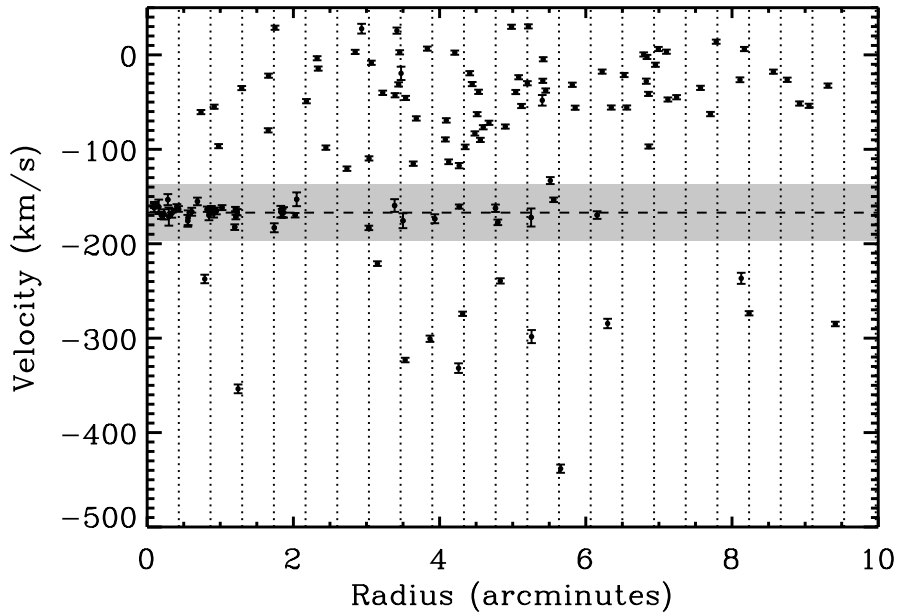}
\caption{Line--of--sight velocities versus distance from the center of
  Seg\,3 for our entire spectroscopic sample.  Vertical dotted lines
  represent radial steps in increments of $r_{1/2} = 26''$ (derived in 
  Section \ref{ssec:struct}).  We detect a
  clear overdensity of points within $v\sim-167\pm30$ km s$^{-1}$ 
  (highlighted in grey),  which we associate with Seg\,3.  The bulk of 
  Milky Way stars lie in the velocity range $\sim -125$ to 25 km s$^{-1}$, more 
  than 40 km s$^{-1}$ away from the velocities of Seg\,3.}
\label{fig:vdist_all}
\end{figure*}

We combine our photometric and spectroscopic data to determine probable members of Seg\,3.  In Figure 
\ref{fig:vdist_all} we present the line--of--sight velocities for our full spectroscopic sample.  We detect a clear 
overdensity of line--of--sight velocities which begin at the center of Seg\,3, and continue out to $\sim14$ times the 
half--light radius of the system ($26''$, see Section \ref{ssec:struct}).  These stars exhibit an average systemic velocity 
of 167 \kms, and are offset by $\sim150$ \kms\ from the center of the Milky Way distribution at this position.  We 
hypothesize that stars in our sample within $167\pm30$ \kms\ are candidate members of Seg\,3.  This velocity range 
is chosen to be wide enough to allow for the inclusion of extra-tidal stars, yet small enough to prevent the inclusion 
of significant numbers of Milky Way stars.  We note, however, that our results are not sensitive to small changes in 
the size of this velocity window.

\begin{figure*}[t!]
\centering
\includegraphics[clip=true, trim=2.2cm 12.1cm 2.1cm 3.cm,width=12cm]{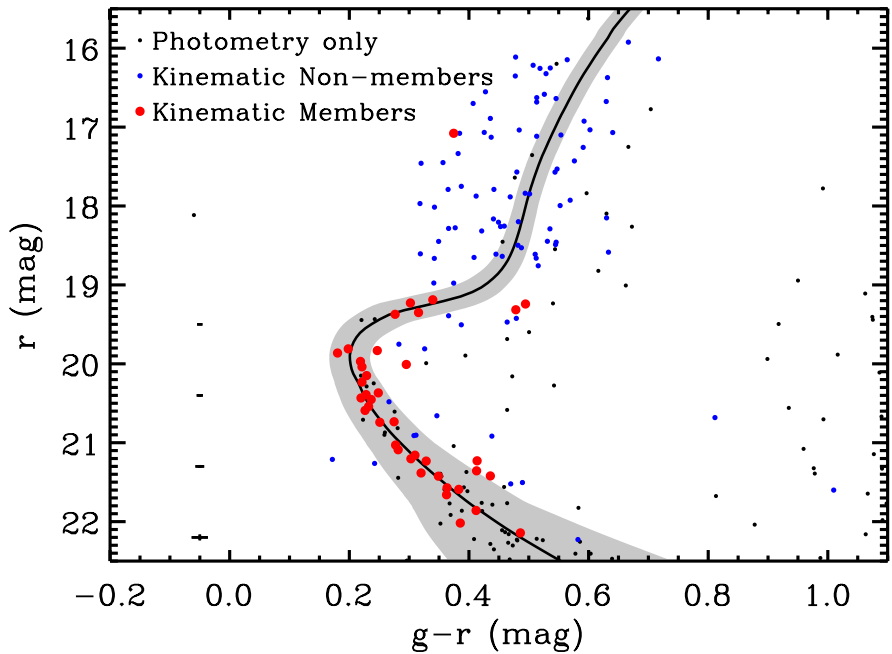}
\caption{The $r$ versus $g-r$ color--magnitude diagram of Segue 3.
  Small black points are stars within $5r_{1/2}$ of the center of Seg\,3 for
  which only IMACS photometric data is available.  Plotted at $g-r =
  -0.05$ are the average photometric uncertainties as a function of
  magnitude.  Small blue (large red) points indicate stars with measured line--of--sight
  velocities that are outside (within) 30 km s$^{-1}$ from the systemic velocity of Seg\,3.  The
  black curve represents the isochrone of a stellar population with an
  age of 12 Gyr and [Fe/H] $= -1.7$ \citep{dotter08}.  We impose a color--magnitude
  selection indicated by the grey band, which roughly
  corresponds to $5\sigma$ photometric uncertainties about the
  isochrone.  Our final list of candidate members for Seg\,3 consists of
  all stars which pass our kinematic and photometric criteria (large
  red points, within grey band).}
\label{fig:cmd}
\end{figure*}

In Figure \ref{fig:cmd}, we present the $r$ versus $g-r$ color--magnitude diagram (CMD) of stars within $5r_{1/2}$ 
of the center of Seg\,3, using our IMACS photometry.  Also plotted in the CMD are stars that are in our spectroscopic 
sample but lie at larger radii.  We find a distinct main sequence of stars associated with Seg\,3, with similar 
photometric properties as those found by \citet{belokurov10a}.  An eyeball fit suggests that this main sequence of 
stars is well described by an isochrone with an age of 12 Gyr and [Fe/H] $=-1.7$ \citep{dotter08}, shifted by a 
distance modulus of 16.15.  In Section \ref{ssec:isofit}, we consider the quality of fits for isochrones with different ages, 
metallicities, and distance moduli.  While these may differ from the isochrone used here, we note selection of our 
member sample will not depend on small deviations in isochrone properties, since we select members using a wide 
$\sim5\sigma$ CMD window.

\begin{figure*}
\centering
\includegraphics[clip=true, trim=1cm 12.1cm 3.5cm 0.cm,width=12cm]{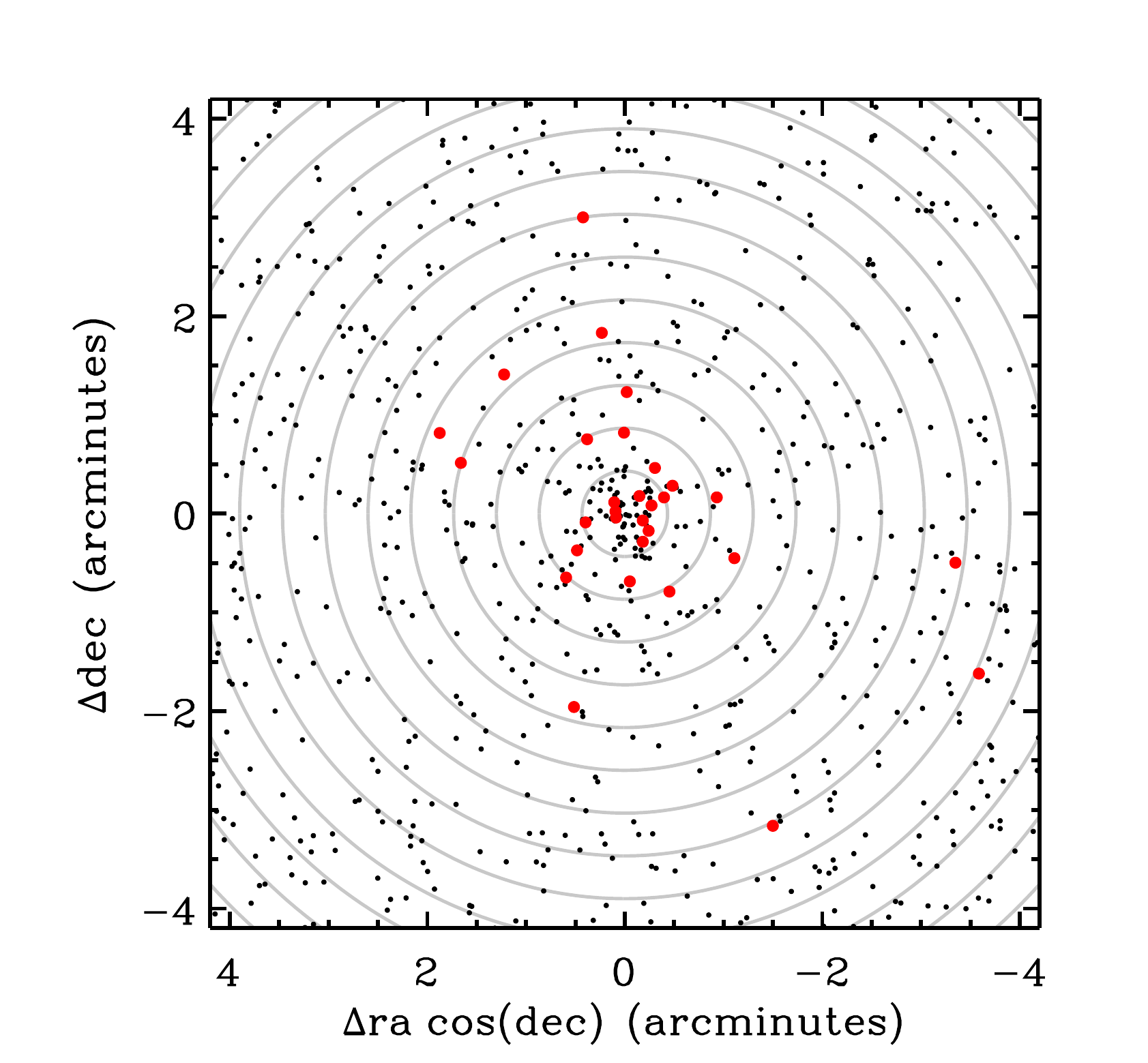}
 \caption{The spatial distribution of our spectroscopically confirmed 
 members of Seg\,3.  Plotted in black are the locations of stars with color 
 $g-r<1.0$ for which we do not have spectroscopic coverage.  Overplotted 
 in red are the locations of identified Seg\,3 members.  Grey contours 
 indicate increasing steps in the half--light radius of Seg\,3.  The apparently 
 irregular distribution of red points reflects the non-uniform spectroscopic 
 coverage, rather than a true irregularity in the underlying distribution of 
 member stars.}
\label{fig:spatial}
\end{figure*}

We impose two selection criteria for probable Seg\,3 members.  First, we kinematically select stars that have 
line--of--sight velocities within $167\pm30$ \kms.  Second, we require Seg\,3 members to pass a photometric cut 
(grey area, Figure \ref{fig:cmd}) which corresponds approximately to the $5\sigma$ average photometric uncertainty 
region around our fiducial isochrone.  In total, we select 32 likely members of Seg\,3 and in Figure \ref{fig:spatial} we 
present the spatial distribution of the members.

\subsection{Foreground Contamination}
\label{ssec_contam}

Examining our sample of Seg\,3 member stars, we find a number of stars which lie at large distances (up to 
$14r_{1/2}$) from the center of Seg\,3.  We consider the possibility that these stars, and those at smaller radii, might 
instead be foreground Milky Way contaminants.  To estimate the degree of foreground contamination in our member 
sample, we utilize the Besancon model of the Milky Way \citep{robin03}.  The estimate does not include possible 
contamination in our Seg\,3 member sample from halo substructures or from unbound stars that are physically 
associated with Seg\,3.

For our estimation we select all Besancon stars that are within one degree of the center of Seg\,3, and are within a 
$10\sigma$ average photometric uncertainty region about our fiducial isochrone (a factor 2 larger than for our 
Seg\,3 members).  Of these stars, we find the number of stars that meet our above kinematic criteria is $3.66 \pm 
0.18$\,\% of those that do not.  In our spectroscopic sample we find 11 stars that satisfy CMD criteria but are not 
within our kinematic window, implying an average of $0.40\pm0.02$ stars are contaminating our sample of Seg\,3 
members.  From this average, we quantify the frequency of larger numbers of contaminants using Poisson statistics.  
We find $N=\{1,2,3,4\}$ contaminants occur with a frequency of $\{26.2,5.52,0.78,0.06\}$\,\%.  Thus, we conclude 
that the likely number of Milky Way field halo contaminants is $\le 2$ at $\sim95\%$ confidence.

%
%
\section{Photometric Analysis of Segue 3}
\label{sec:members}

\subsection{Stellar Population}
\label{ssec:isofit}

We now use our spectroscopically selected members to derive the age, [Fe/H], and distance of Seg\,3, using a 
maximum likelihood method which closely follows that described by \citet{frayn02a}.  For the procedure, a suite of 
isochrones are fit to a sample of stars, assigning to each a bivariate Gaussian probability function whose variance 
is set by the associated photometric errors.  We apply this analysis using all stars in our photometric data within a 
radius $39''$ from the object's center \citep[$r_{1/2}$ from ][]{belokurov10a}, to which we add the 17 spectroscopically 
confirmed member stars that lie outside that area, resulting in a total sample of 125 stars.

For a given isochrone $i$ we compute the likelihood 

\begin{eqnarray}
\mathcal{L}_i=\prod_j p(\{g,g-r\}_j|i,\{g,g-r\}_{ij},{\rm dm}_{0,i}) ,
\end{eqnarray}

\noindent where  

\footnotesize

\begin{eqnarray}
\label{eqn:eachlike}
&&p(\{g,g-r\}_j|i,\{g,g-r\}_{ij},{\rm dm}_{0,i})=\frac{1}{2\pi\sigma_{g_j}\sigma_{(g-r)_j}} \times  \\ 
&&\exp\left(-\frac{1}{2}\left[\left(\frac{g_j-(g_{ij}+{\rm dm}_{0,i})}{\sigma_{g_j}}\right)^2 + \left(\frac{(g-r)_j-(g-r)_{ij}}{\sigma_{(g-r)_j}}\right)^2 \right]\right)\nonumber .
\end{eqnarray}

\normalsize

\noindent For each star $j$, $\{g,g-r\}_{ij}, {\rm dm}_{0,i}$ are the magnitude, color, and de-reddened distance 
modulus values for isochrone $i$ that maximize the likelihood of the entire data set $\{g,g-r\}_{j}$ in Equation 
\ref{eqn:eachlike}.  We take an approximate solution to finding the values of $\{g,g-r\}_{ij}$ and $\rm dm_{0,i}$ by 
searching over a series of fine steps in $g,g-r,$ and $\rm dm_0$ values for each isochrone.  Input isochrones are 
supplied by the Dartmouth library \citep{dotter08}, and interpolated so $g,g-r$ values step by 0.001 mag in the 2D 
color--magnitude space.  The distance modulus $\rm dm_0=m-M$ is sampled over a range of $15.0 < \rm dm_0 < 17.0$ 
in steps of 0.01 mag.

\begin{figure*}[t]\centering
\includegraphics[clip=true,trim=2.2cm 11.1cm 1.6cm 5.cm, width=12cm]{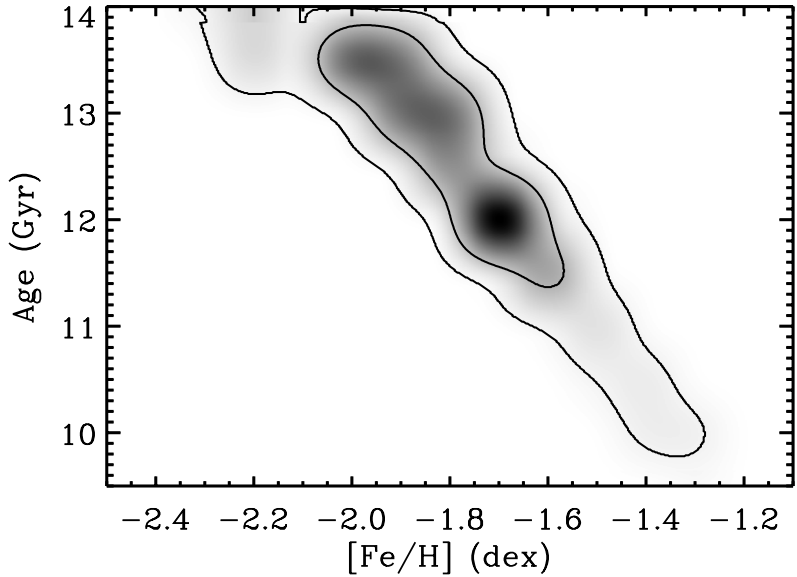}
\caption{Smoothed maximum-likelihood joint probability density in age-metallicity space for all Segue 3 stars within a
radius of $39''$, combined with 17 spectroscopically confirmed stars that lie outside that area.
Contour lines show the 1 and 2$\sigma$ confidence levels. The diagonal flow of the contour lines reflect the
age-metallicity degeneracy inherent to such an isochrone fitting procedure. The 1D marginalized parameters 
around the best fit are:  
age $=12.0^{+1.5}_{-0.4}$\,Gyr, [Fe/H]$=-1.7^{+0.07}_{-0.27}$\,dex, $m-M=16.14\pm0.09$\,mag.
}
\label{fig:isolike}
\end{figure*}

We calculate the maximum likelihood values $\mathcal{L}_i$ over a grid of isochrones, covering an age range from 8 
to 14\,Gyr and metallicity range $-2.5\leq$ [Fe/H] $\leq-0.9$\,dex.  Grid steps are 0.5 Gyr in age, and 0.1 dex in [Fe/H].  
With a grid of $\mathcal{L}_i$ values, we can locate the most likely value and compute confidence intervals by 
interpolating between grid points.  In addition to this interpolation, we smooth the likelihood values over $\sim2$ grid 
points in order to provide a more conservative estimate of parameter uncertainties.  In Figure \ref{fig:isolike}, we 
present the relative density of likelihood values for the sample described above.  We find the isochrone with the 
highest probability has an age of 12.0 Gyr and [Fe/H] $= -1.7$, with 68\% and 95\% confidence contours presented in 
the figure.  The marginalized uncertainties about this most probable location correspond to an age of 
$12.0^{+1.5}_{-0.4}$\,Gyr, a metallicity of [Fe/H]$=-1.7^{+0.07}_{-0.27}$\,dex, and a distance modulus of 
$\rm dm_0=16.14\pm0.09$\,mag ($d = 16.9\pm0.7$ kpc).  We assume a distance of 17 kpc in the calculation of 
physical size and absolute magnitude in Section 4.2.

To assess how much the inclusion of the 17 spectroscopic members outside of $39''$ has influenced the above results, 
we repeat the above computation using only the 32 spectroscopic members found above.  We find parameters derived 
for this subsample are consistent with the larger sample but have larger uncertainties, with age $=13.5^{+1.5}_{-1.3}$\,Gyr,
[Fe/H] $=-1.9^{+0.20}_{-0.39}$\,dex, and $\rm dm_0=16.08\pm 0.13$\,mag ($d = 16.4\pm1.0$ kpc).

\subsection{Structural Parameters}
\label{ssec:struct}

We present a new determination of the structural parameters of Seg\,3, 
using the photometry presented in Section \ref{ssec:phot}. For our 
analysis we follow the maximum likelihood method of \citet{martin08b}, 
as described in \citet{munoz10a}.  This method starts by assuming an 
analytic surface density profile, and then fits the profile parameters using all 
stars meeting CMD criteria, thus avoiding the need to bin or smooth data.  
We fit structural parameters for Seg\,3 using two choices of density profiles 
commonly used to describe the light distribution in ultra-faint systems, 
an exponential and a Plummer \citep{Plummer11} profile. In both 
cases, the parameters we calculate are the scale length of the system, 
the coordinates of its center, its ellipticity, position angle and the 
foreground/background stellar density.  To estimate parameter 
uncertainties, we carry out a bootstrap analysis using $10^4$ 
realizations of the photometric data.  

\begin{figure*}[t]\centering
\includegraphics[clip=true,trim=0cm 0cm 0cm 0.cm, width=8cm]{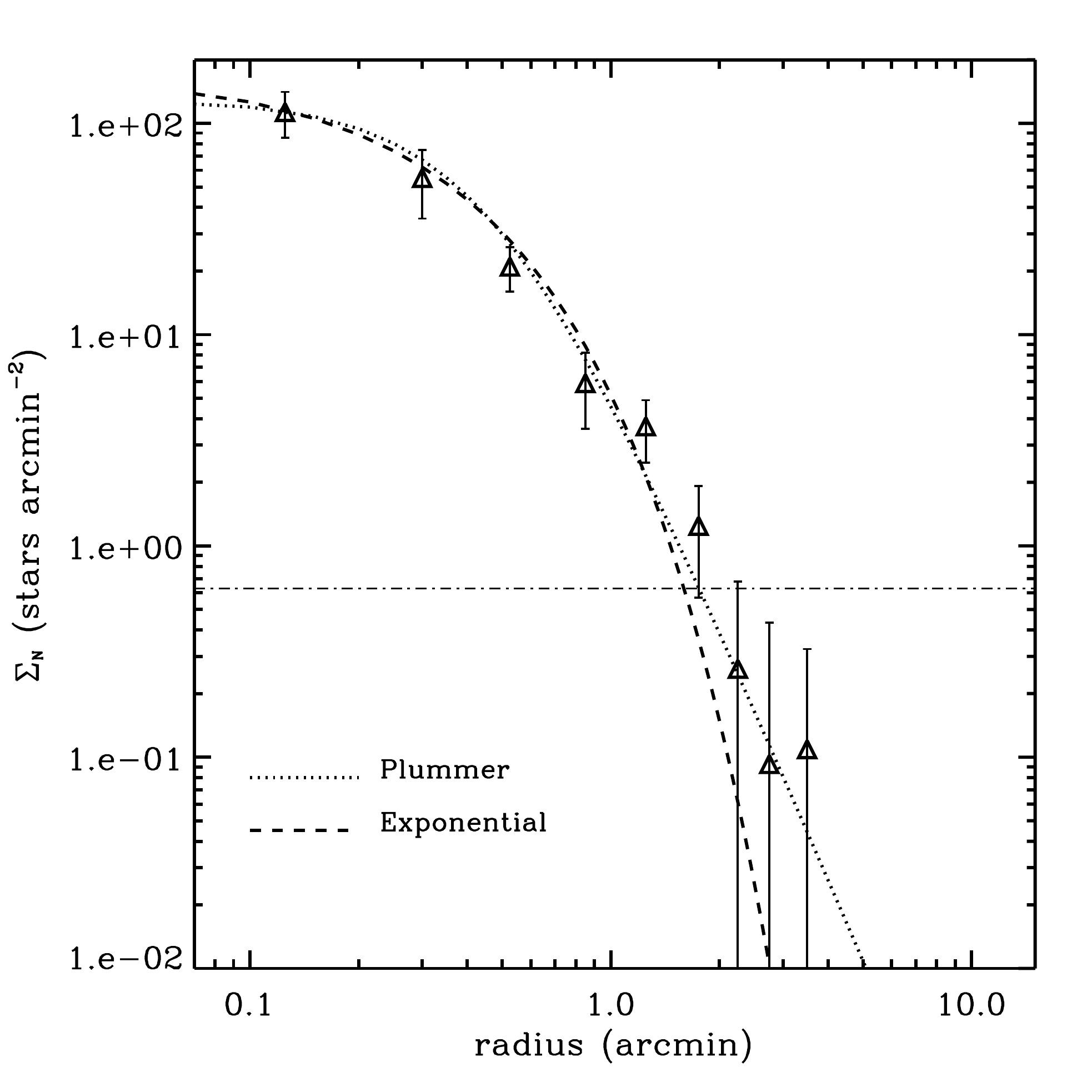}
\includegraphics[clip=true,trim=0cm 0.1cm 0cm 1.4cm, width=8.6cm]{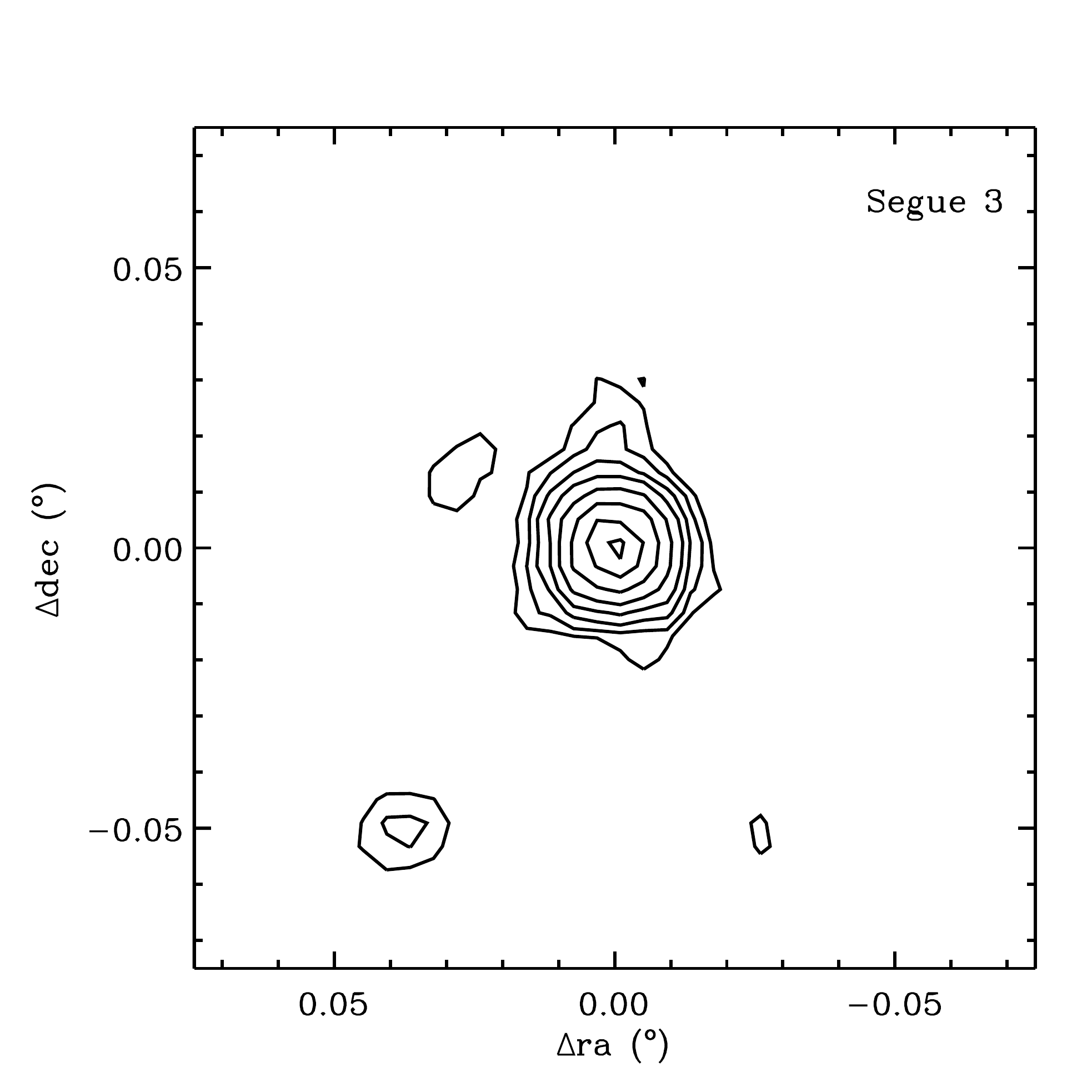}
\caption{\textit{Left:} 1D surface density profile of Seg\,3 for a Plummer and 
exponential profile, both of which provide acceptable fits to the data.  The dot-dashed 
line represents the surface density of background stars inferred by our analysis.  
\textit{Right:} 2D surface brightness contours for Seg\,3.  Contour levels are for 
$\{3,5,8,12,18,26,36,44\}\sigma$ fainter than the peak surface brightness, where 
$5\sigma$ is 2.35 mag arcmin$^{-1}$ fainter than the peak value of $\mu_{0,V}=$ 24.1 
mag arcmin$^{-1}$ (assuming a Plummer profile).  We find the stellar distribution is 
approximately circularly symmetric until it falls to levels comparable the background.
}
\label{fig:struct}
\end{figure*}

Before proceeding with our analysis of the structure of Seg\,3, we must 
carefully consider which photometric data to use.  We note in Section 
\ref{ssec:phot} that the 90\% completeness limit of our data is 
at a magnitude of $\sim23.8$ for $g$ and $r$ filters.  However, due to the 
increased contamination of unresolved galaxies beyond 
a magnitude of $r\sim23$, we conservatively limit our structural 
analysis to stars brighter than $r\sim22.5$.  In Table \ref{tab:struct} we 
present the results for both Plummer and exponential density 
profiles.  We find structural parameters very similar to those of 
\citet{belokurov10a}, with most results within $1\sigma$ of values 
previously reported.  A notable exception, however, is our determination 
of the half--light radius of $r_{1/2} = 28''\pm8''$ and $26''\pm5''$, or $2.2\pm0.7$ pc
and $2.1\pm0.4$ pc for an 
exponential and Plummer profile, respectively.  While within 
$\sim1.5\sigma$ from the value of \citeauthor{belokurov10a}, these are $\sim30\%$ 
smaller than previously found.
We attribute such discrepancies to our higher quality photometric data, and to the 
different analysis techniques used here.
In Figure \ref{fig:struct} we present the 1D surface density profile and 2D surface 
brightness contours for Seg\,3.  Both the Plummer and exponential profiles are found 
to provide good fits to the data, while two--dimensional contours reveal 
a circularly symmetric profile, with little evidence for strong tidal 
distortion.

In addition to structural properties, we estimate the absolute magnitude of 
Seg\,3 following the method described in \citet{munoz10a}.  Previous estimates 
of the absolute magnitude by \citet{belokurov10a} reported a value of 
$M_{V}= -1.2$ for Seg\,3.  Such a low magnitude makes traditional methods 
for calculating total luminosities, such as adding individual stellar fluxes, too 
sensitive to the inclusion (or exclusion) of potential members (outliers) \citep[e.g.,][]{walsh08a,
martin08b,munoz10a}.  To alleviate issues related to low number statistics, 
the method used here relies solely on the total number of stars that belong 
to the satellite and not on their individual magnitudes.  

In short, our method estimates the absolute magnitude by integrating a 
theoretical luminosity function, using the above structural analysis.  Following 
the results of Section \ref{ssec:isofit}, we use a luminosity function for a $12$\,Gyr 
stellar population with a [Fe/H]$=-1.7$ from \citet{dotter08}, assuming a 
Salpeter IMF.  We then integrate this luminosity function down to the magnitude limit 
of $r=22.5$, and scale the results to match the number of stars found by our 
structural analysis. 
Our final estimate of the absolute magnitude, therefore, is the integral of 
this scaled luminosity function, integrated down to the limiting mass of the Salpeter IMF 
($0.1 M_\Sun$).  Like our structural parameters, the uncertainty of the 
absolute magnitude is estimated by bootstrapping the data $10^4$ times. 
This method yields $M_{V}=-0.06\pm0.78$ or
$L_{V}=90^{+95}_{-56}$\,L$_{\sun}$ for an exponential profile. 
If we use a Plummer profile instead, we obtain $M_{V}=-0.04\pm0.78$ 
which translates into $L_{V}=89^{+93}_{-45}$\,L$_{\sun}$.  While the uncertainties 
are significant, these values make Segue 3 the lowest luminosity stellar 
system known to date.  This tiny luminosity combined with the small size of Seg\,3 yields 
a brighter central surface brightness ($\mu_{0,V} = 23.90^{+1.0}_{-0.8}$ for an exponential 
profile, $\mu_{0,V} = 24.10^{+1.0}_{-0.8}$ for a Plummer profile) than that of the most diffuse 
Milky Way companions.  

We note, 
in fact, that our values for the luminosity and surface brightness of Seg\,3 may 
be overestimated if stars fainter than the magnitude limit of our observations have 
be preferentially lost due to relaxation (see Section \ref{sec:dynamical}).  While it 
is unclear whether relaxation is significant in Seg\,3, such an effect would result 
in only a small change in our estimates, since low mass stars do not contribute 
much to the total luminosity.  For the remainder of the analysis presented here, we adopt the structural 
parameters and absolute magnitude obtained from using a Plummer 
profile above.

%
%
\section{Segue 3: A Probable Star Cluster}

We assess the nature of Segue 3 through several independent lines of evidence.  First, in Section \ref{ssec:vdist} we examine 
the velocity distribution of Seg\,3 spectroscopic members.  In Section \ref{ssec:agemetal}, 
we consider possible spreads in the distribution of ages and metallicities 
of Seg\,3 spectroscopic member stars.  Finally, in Section \ref{ssec:size} we examine 
the size and luminosity of Seg\,3 relative to other known Milky Way satellites.  
As discussed below, we conclude that all three indictors agree with our null 
hypothesis: Segue 3 is an old, extremely low luminosity star cluster in the 
halo of the Milky Way. 

\subsection{Velocity Distribution}
\label{ssec:vdist}

\begin{figure*}
\centering
\includegraphics[clip=true, trim=1.2cm 12.1cm 1.6cm 2.cm,width=12cm]{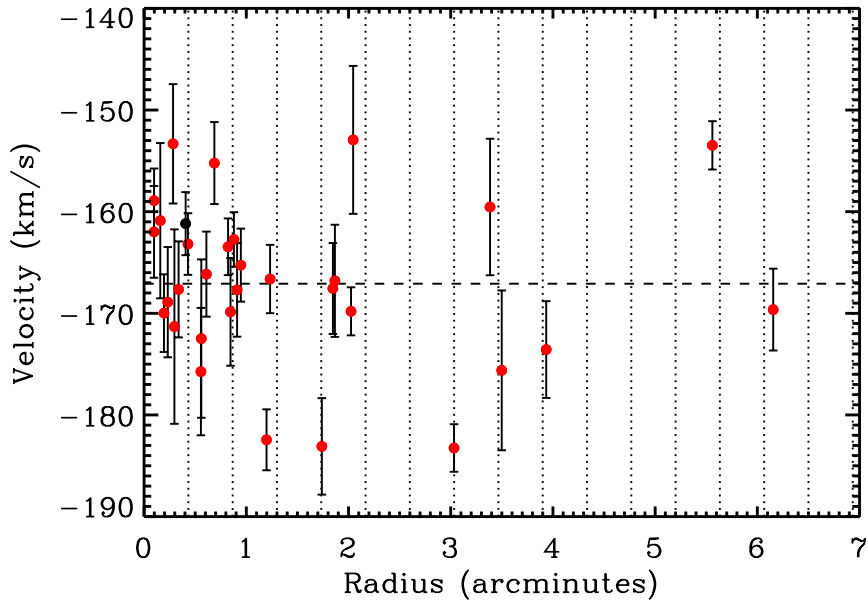}
 \caption{Line--of--sight velocities of members versus distance from the center of Seg\,3.
 Plotted in black is the combined velocity for the binary star found in our repeat 
 measurements (described in Section \ref{ssec:spect}). The dashed line indicates the 
 systemic velocity at $-167$ \kms, while the dotted lines indicate increasing 
 steps in $r_{1/2}$ from the center.}
\label{fig:vdist}
\end{figure*}

We present in Figure \ref{fig:vdist} the distribution of velocities for 
Seg\,3 spectroscopic members.  We find the majority of the members (66\%) lie at small 
projected radii ($\le 3r_{1/2}$), with the remainder up to 14 half--light radii 
from the center of Seg\,3.   We 
consider these eleven spatially outlying stars stars, as well as other lines of 
evidence, as possible signs of stellar mass loss in Seg\,3 (discussed below in 
Section \ref{sec:dynamical}).  

To obtain a reasonable estimate of the velocity dispersion (and hence 
dynamical mass) of Seg\,3, we choose to focus on the velocity 
measurements within $3r_{1/2}$ only, and exclude the binary star 
identified by repeat velocity measurements (ID 9, Table \ref{tab:mem}).  
This choice excludes distant candidate member stars potentially unbound to Seg\,3, and 
reduces the amount of foreground contamination, since foreground stars are more likely 
at larger radii.  In estimating the velocity dispersion, we 
use the maximum likelihood technique described by \citet{walker06a}.  

If we naively include all 20 stars within $3r_{1/2}$, we infer a velocity dispersion of 
$5.3\pm1.3$ \kms\ for Seg\,3.  However, this relatively large dispersion is 
primarily driven by the presence of a single outlying star (ID 20, Table 
\ref{tab:mem}), with a measured velocity of $-182\pm3$ \kms, $\sim5\sigma$ 
away from the systemic velocity of Seg\,3.  Omitting this star, our estimate of the 
velocity dispersion plummets to $1.2\pm2.6$ \kms, a value consistent with zero.

Inclusion of the outlying star has a profound consequence on the nature 
of Seg\,3.  If included, we must necessarily infer the presence of 
significant amounts of dark matter in Seg\,3, since the velocity dispersion would 
imply a mass--to--light ratio of $645^{+1286}_{-442}$ within $r_{1/2}$.\footnote{Using the 
mass estimator of \citet{wolf10a}, assuming a constant velocity dispersion within $3r_{1/2}$.}  
If omitted, the velocity distribution allows very small 
values of the dispersion, giving dynamical masses consistent with stellar 
material alone.  We must, therefore, carefully consider whether to omit 
the outlying star.  

We use Monte Carlo simulations to test the hypothesis that the outlying 
star is revealing the true velocity dispersion, rather than being a genuine 
outlier from the velocity distribution.  First, we assume a value for the true, 
intrinsic velocity dispersion of Seg\,3 ranging from 0 to 10 \kms.  We then 
repeatedly generate a sample of 20 stars with velocities drawn randomly 
from a Gaussian corresponding to the intrinsic dispersion, convolved with 
our DEIMOS measurement uncertainties.  Of these samples, we take 
those which have a dispersion between $5.3\pm1.3$ \kms\ and ask how frequently 
the dispersion is reduced to below $3.8$ \kms, once we omit the largest 
outlier.  Our simulations indicate the highest 
frequency occurs at an intrinsic dispersion of 5.8 \kms, at a rate of 3.8\%.  
Therefore, at $>95\%$ confidence, our simulations indicate the outlying 
star within $3r_{1/2}$ is not due to low--number statistics, but is instead a 
genuine outlier from the distribution.  With this conclusion, we assume it 
is reasonable to omit the outlying star, and infer a velocity dispersion of 
$1.2\pm2.6$ \kms\ within $3r_{1/2}$.  By adopting such a velocity 
dispersion, we can infer a mass--to--light ratio of $33^{+156}_{-144}$ 
within $r_{1/2}$\footnotemark[6].  While inconclusive, from this 
mass--to--light ratio alone there is no compelling evidence for significant 
dark matter content in Segue 3.

\subsection{Age and Metallicity}
\label{ssec:agemetal}

Properties of the stellar populations of Milky Way satellites have proven 
to be a useful indicator for the presence of an underlying dark matter 
halo.  In short, the deeper potential well provided by non-baryonic material 
enables star formation processes to withstand feedback effects from (e.g.,)
supernovae, facilitating more extended episodes of star formation.  This 
effect manifests itself in the presence of a range of metallicities amongst 
stars, since metals from earlier generations of stars may be incorporated 
into subsequent ones.  Ultra--faint satellites have demonstrated this 
phenomenon, showing internal [Fe/H] spreads up to 0.5 dex or more 
\citep{simon07a,kirby08a,kirby11a,simon11a,willman11a}.

\begin{figure*}[t!]
\centering
\includegraphics[clip=true, trim=2.2cm 12.1cm 1.3cm 2.cm,width=12cm]{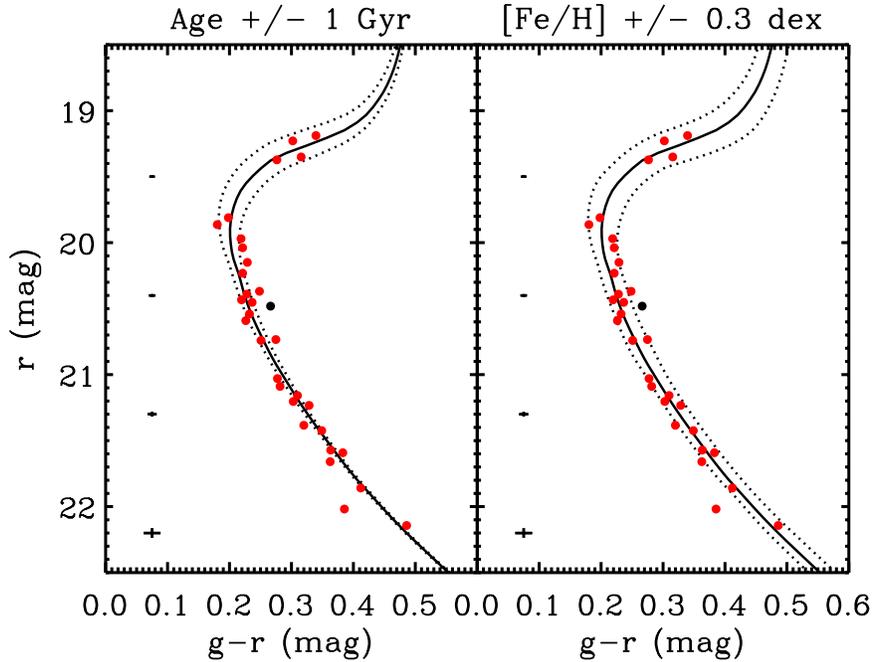}
\caption{The color--magnitude diagram from Figure \ref{fig:cmd},
  showing only member stars of Seg\,3.  The black point indicates 
  the color and magnitude of the binary star found with repeat 
  velocity measurements (see Section \ref{ssec:spect}).  Average photometric
  uncertainties are plotted on the left in each panel.  The solid,
  black curve represents our fiducial isochrone from Figure
  \ref{fig:cmd}.  In the left panel, dotted curves show the isochrones
  for a stellar population which are older by $\pm1$ Gyr.  In the
  right panel, dotted curves show the isochrones for a stellar
  population with [Fe/H] values which differ by $\pm0.3$ dex.  Within
  the photometric uncertainties, the members of Seg\,3 are consistent
  with a spread in age ([Fe/H]) less than 1 Gyr (0.3 dex).}
\label{fig:cmd_agemetal}
\end{figure*}

In Figure \ref{fig:cmd_agemetal}, we present the effect of varying the 
age and metallicity of isochrones from our fiducial values 
of 12 Gyr and [Fe/H] $=-1.7$.  In the right panel, we see that the 
members of Seg\,3 are consistent with a spread in [Fe/H] $<0.3$ dex, 
within the photometric errors.  Such dispersions in [Fe/H] are 
smaller than values typically found in ultra--faint systems 
\citep{ frebel10a,kirby11a}, indicating the stellar population of Seg\,3 
may be quite different from dark matter dominated satellites.

In addition to internal spreads in metallicity, the average value of [Fe/H] 
in stars can also be used to test whether Seg\,3 may contain significant dark matter.  
\citet{kirby11a} show a strong correlation between the 
total luminosity and [Fe/H] for Milky Way dwarf galaxies, such that the faintest satellites are the most 
metal poor.  Extrapolating the relation from \citeauthor{kirby11a}, we find 
Seg\,3, with a absolute magnitude of $M_V\sim0.0$, should exhibit a mean 
[Fe/H] $<-3.0$, far lower than the $-1.7^{+0.07}_{-0.27}$\,dex we observe.  For comparison, Segue 
1 ($M_V\sim-1.5$) has a value of [Fe/H] $\sim-2.5$ \citep{simon11a}.   For this to be a meaningful 
comparison, we assume that Seg\,3 would not have undergone massive stellar loss if it contains 
significant dark matter.

The low spread and (relatively) high mean of [Fe/H] values inferred 
from isochrone comparisons indicates that the metallicity characteristics of Seg\,3 are 
distinct from that of dark matter dominated ultra--faint 
satellites.  While enticing, several caveats exist.  First, our sample of 
Seg\,3 members is small, totaling 32 stars.  More extensive samples 
might show larger scatter than seen here.  In addition, significant 
uncertainties exist in the values of theoretical isochrones, especially 
under the correlated effects of varying age and [Fe/H] values.  
We have only explored one such set of isochrones here \citep{dotter08}.
To avoid such uncertainties, a definitive measurement of the metallicity 
would require spectroscopic measurements of [Fe/H] values.  For our current data, 
this would require a calibration of the relationship between [Fe/H] and 
the Ca II triplet for main sequence stars. Since no such relation exists, we do not 
attempt to relate the values of the Ca II triplet in our spectra to [Fe/H].  Nevertheless,
we note that the scatter in Ca II equivalent width is consistent with no 
abundance spread.

\subsection{Size and Luminosity}
\label{ssec:size}

\begin{figure*}[t!]
\centering
\includegraphics[clip=true,trim=0.5cm 0.7cm -0.2cm 0.5cm,width=12cm]{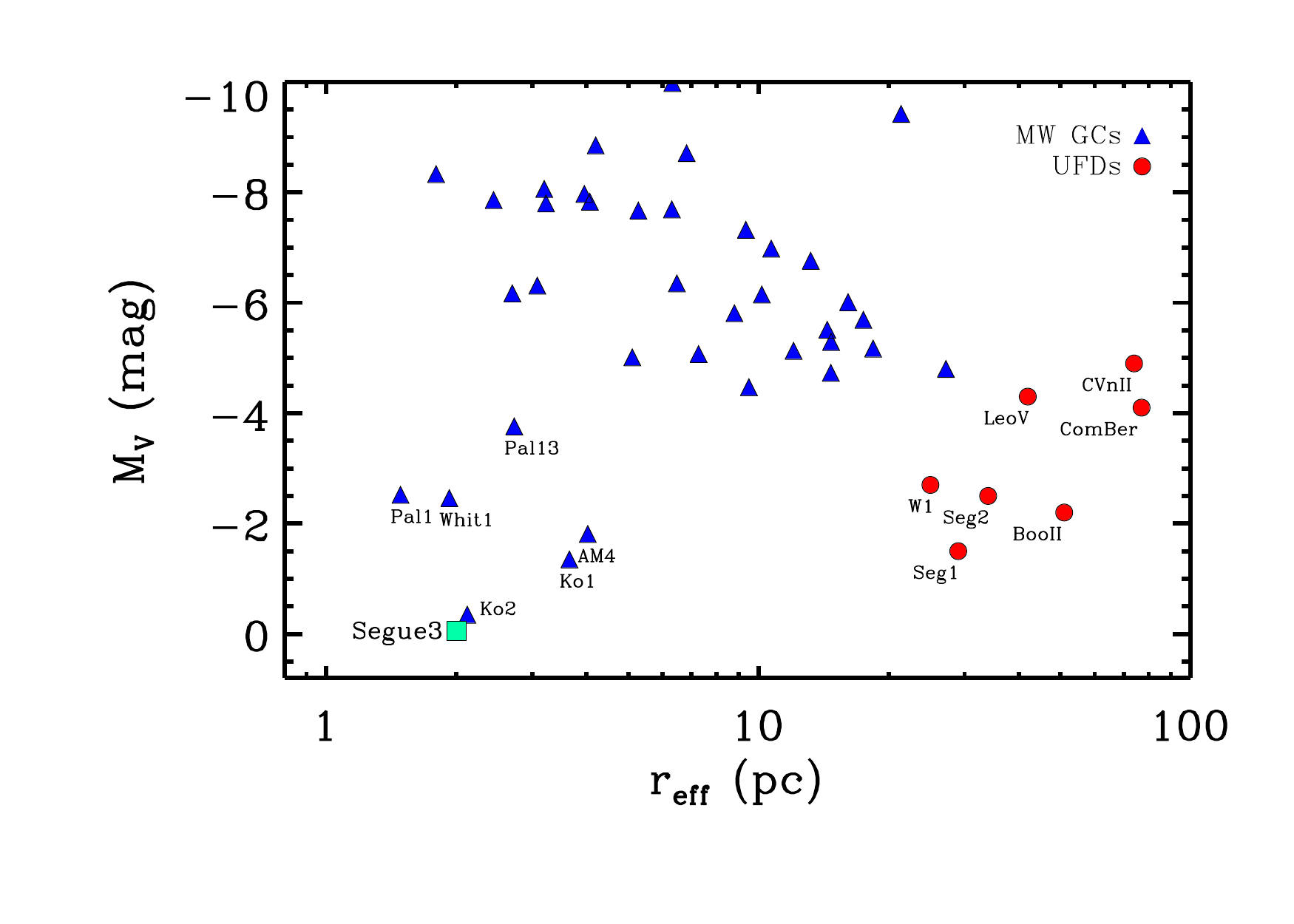}
\caption{The size and luminosity of Milky Way halo objects ($>15$ kpc), presented 
as the effective radius and absolute V--band magnitude, respectively.  At fixed 
luminosity, objects with significant dark matter (ultra--faint and dwarf spheroidal 
galaxies) are larger in size by up to an order of magnitude.  Moreover, faint 
($M_{\rm V}<-4$) star clusters show effective radii $<10$ pc in size.}
\label{fig:size}
\end{figure*}

A final diagnostic tool for discriminating star clusters from satellites with dark matter halos 
is the relationship between size 
and luminosity of Milky Way objects.  In Figure \ref{fig:size}, we show the 
sizes and luminosities of Milky Way halo objects (distances $>15$ kpc).  Of 
these systems we consider two classes, stellar systems with dark matter 
(dwarf spheriodal, ultra--faint dwarf galaxies) and those without dark matter 
(e.g., globular clusters).  A clear trend is present between the two:  at fixed luminosity, 
objects with dark matter are consistently larger in size, often by up to an order of 
magnitude, than objects classified as star clusters.   Segue 3's size and luminosity are far 
smaller than those of objects known to be dominated by dark matter.  This difference 
provides additional circumstantial evidence that Seg\,3 is a star cluster.  Such 
a conclusion is further supported by theoretical arguments, which expect low 
mass dwarf galaxies to be significantly larger in size \citep[e.g.,][]{bullock10} than Seg\,3.  

Under the three lines of evidence presented in 
this section, we conclude that Segue 3 is likely a star cluster with little to no 
dark matter content.  This hypothesis is further supported by tentative signs 
for stellar mass loss (see below), which has also been seen in the majority of 
small and faint stellar clusters. 

%
%
\section{The Dynamical State of Segue 3}
\label{sec:dynamical}

Discussed in Section \ref{ssec:vdist}, the distribution of velocities
within $3r_{1/2}$ of the center of Segue 3 imply a velocity dispersion
of $1.2\pm2.6$ \kms.  The distribution
of these velocities are consistent with being drawn from a
Gaussian distribution, suggesting that Seg\,3 is bound within
$3r_{1/2}$.  

Although its not possible to draw a robust conclusion about the extended stellar 
distribution of this extremely low luminosity system, if we consider Seg\,3 members 
across all radii, then we find tentative evidence for mass loss in Seg\,3.  The fact 
that 1/3 of our candidate member stars lie outside of three half-light radii hints that 
there may be an excess of Seg\,3 member stars at large distances, as expected if it 
is currently undergoing stellar mass-loss.  We attempted to quantify this excess (or 
lack thereof) by correcting the surface density profile of spectroscopic members for 
target efficiency and comparing with expectations from a pure Plummer model.  
However,  uncertainty owing to small number statistics in both our characterization 
of target efficiency (as a function of position), and in our predicted number of stars at 
large radii, was too great to make a meaningful comparison with observations. 

Of the thirteen stars at radii $>2.5r_{1/2}$,  four  (IDs 20, 22, 27, and 31 in Table
\ref{tab:mem}) lie at velocities more than 10 \kms\ away from the systemic velocity of $-167.1\pm1.5$\kms. 
Three of the four stars have 
velocities which are each $\sim 16$\kms\ below the systemic velocity of 
Seg\,3, while the fourth has a velocity $\sim 14$\kms\ above the systemic. 
In a sample of 32 stars, the likelihood of finding four such stars in a tidally 
isolated system is small.  Performing similar Monte Carlo tests as in Section
\ref{ssec:vdist}, we determine these outlying stars are not due to
small sample size at $>99\%$ confidence.  These stars may be stars lost by Seg\,3 
and heated up by a dynamical interaction, or they may be contaminants from the field 
halo or from the possible proximity 
of Seg\,3 to to another known halo overdensity (see Section \ref{ssec:HercAq}).
However, simple estimates presented in Sections \ref{ssec_contam} and 
\ref{ssec:HercAq} indicate the chance of all four stars being 
contaminants is excluded at $>99\%$ confidence.  

While such simple estimates may underestimate the degree of contamination
of our member sample, 
mass loss in Seg\,3 is nevertheless supported from dynamical arguments.  
First, we consider the possibility of tidal stripping in Seg\,3 by estimating 
the Jacobi radius $r_{\rm J}$ of the system.  Since $r_{\rm J}\propto 
M_{\rm MW}^{-1/3}$, we conservatively overestimate $r_{\rm J}$ by 
assuming a Milky Way circular velocity of only 180 \kms at the Galactocentric distance 
of Seg\,3 (10 kpc).  This yields a value of $r_{\rm J}
=10.4$ pc, $\sim5$ times the half--light radius of Seg\,3.  Given
that Seg\,3 is likely not on a purely circular orbit, and may be
associated with a local overdensity (see below), the true tidal radius
of Seg\,3 is likely smaller still.  Thus, from these simple estimates,
it seems highly likely that Seg\,3 is undergoing some sort of tidal
stripping.

In addition to tidal stripping, we consider potential evidence for
relaxation in Seg\,3, using the half--mass relaxation time $t_{\rm rh}$ \citep{spitzer71, 
gnedin97}.  Using our measurements of the structural properties of Seg\,3 and assuming 
a Coulomb logarithm of $\ln(\Lambda)=\ln(0.02N)$, we estimate 
$t_{\rm rh}=144^{+153}_{-53}$ Myr.  The timescale for total disruption of a cluster through 
relaxation is of order $10-50 t_{\rm rh}$ \citep[see, e.g.,][]{gnedin99}, implying an 
evaporation time of $\sim 1-7$ Gyr for Seg\,3.  Such evaporation times are 
short relative to the estimated age of the stellar population (12 Gyr), indicating 
Seg\,3 must have experienced significant mass loss in order 
to have survived to present day.  Definitive support for relaxation can be obtained 
if signs of mass segregation are found in the system.  Unfortunately, we are not able to see 
any clear signs of mass segregation, due to the small size and mass range ($\sim0.2\,M_\Sun$) 
of our member sample.  

Interestingly, the mass loss implied by the small tidal radius and small relaxation 
time of Seg\,3 are similar to that found in other faint stellar systems \citep{koposov07b, carraro09a, 
carraro07a,NO10a}.  For instance, \citet{koposov07b} 
estimate Koposov 1 and 2 have $t_{\rm rh}$ $\sim$ 70 and 55 Myr, and 
$r_{\rm J} \sim$ 11 and 9 pc, respectively.  These facts, and the tentative 
associations to large--scale Galactic structures (see below), may indicate a 
common evolutionary history for these systems.

%
%
\section{Proximity to known stellar halo structure}
\label{ssec:HercAq}

Segue 3 lies at a distance of $\sim 17$\ kpc from the Sun, at a location of 
$(l,b)$ = $(69.4,-21.27)$.  This is 
coincident in projection with the recently discovered Hercules--Aquila cloud 
\citep{belokurov07b}.  Early discovery data (SDSS DR5)
presented by \citeauthor{belokurov07b} 
indicated the cloud is located at $l \approx 40^\circ$, and extends
Galactic latitudes as low as $-40$ at $l \sim 50^{\circ}$.  

Arguments for the spatial overlap of Seg\,3 and extended halo substructure have been
strengthened by the analysis of SEGUE/DR7 SDSS data by
\citet{dejong10a}.  Subtracting off a smooth model for the stellar
distribution, \citeauthor{dejong10a} find a number of new halo overdensities.
Of these, one lies at $(l,b,D_{helio}) = (70^\circ, -22^\circ \pm 1,
15\pm1\,{\rm kpc})$, in excellent agreement with the location of
Seg\,3.  While coincident with the Hercules-Aquila Cloud in projection, the distance to stars 
in this newly identified halo structure (and in Seg\,3) are closer than the $>$ 20 kpc distances 
of the Herc-Aquila debris in that direction.  It is presently unclear whether or not these 
structures all have a common origin.  

Although Seg\,3 appears to be coincident with a spatially extended \citet{dejong10a} halo 
overdensity, this halo overdensity isn't necessarily kinematically associated with Seg\,3.  There 
is a dearth of (non-member) stars  with velocities $\pm45$\kms\ from the systemic velocity of 
Seg\,3, suggesting that this structure may not be kinematically associated with Seg\,3. 
As shown in Figure \ref{fig:vdist_all}, we find a significant number
of stars with large negative velocities, more than 40\kms\ lower than
the systemic velocity of Seg\,3.  These stars are not associated with
any other known MW structure, and may be attributed to the \citet{dejong10a} 
$(l,b) = (70^\circ,-22^\circ)$ overdensity.  It is unlikely that these stars are instead Hercules-Aquila 
stars, because their color-magnitude distribution is consistent with a more nearby stellar population 
than the Herc-Aquila cloud in this direction.  

If we assume the stars in this kinematic halo structure are normally
distributed, we find their distribution has mean of $-290$\kms\ and
FWHM of 48\kms.  Under this assumption, we would expect to find an
average of $\sim 0.2$ stars that meet our velocity criterion for
membership in Seg\,3 (cf. Section \ref{sec:members}).  Accounting for
Poisson variance, therefore, we expect this low velocity population of
stars ($<-200$\kms) may be contaminating our Seg\,3 member sample by
at most one star (99\% CL).  Thus, given our contamination estimates for 
both the MW foreground and for this additional kinematic halo population, we can rule out 
contamination from these source as the origin for \textit{all} four outliers in Section 
\ref{sec:dynamical} at $>99\%$ confidence.

Other 
faint stellar systems (like Koposov 1) have also been linked to large 
streams and structures \citep{frinchaboy04,carraro07a,koposov07b,
carraro09a,forbes10a}.  Because of the two distinct halo kinematic populations in our data, 
it is unclear whether Seg\,3 is associated with the \citet{dejong10a} $(l,b) = (70^\circ,-22^\circ)$ 
overdensity.  It seems likely that Seg\,3 is  kinematically offset from this overdensity.  A more extensive 
photometric-kinematic study of the halo debris in this region of sky might be able to reach a more robust 
conclusion about the association, or lack thereof, of Seg\,3 and other halo structures.

%
%
\section{Conclusions}

We have investigated the nature of the faint Milky Way satellite Segue 3 
using new Keck/DEIMOS spectrocopic and Magellan/IMACS photometric 
data.  Using the data we have identified a sample of 32 probable member 
stars for Seg\,3, one of which is definitively identified as a binary system 
through repeated velocity measurements.  We summarize our conclusions 
as follows:

\begin{itemize}

\item New IMACS photometry reveals a distinct main sequence of stars in 
the CMD of Seg\,3, with the majority of these stars lying within two half--light 
radii of the center.  Using maximum likelihood methods which fit isochrones 
from \citet{dotter08}, we infer an average age of $12.0^{+1.5}_{-0.4}$\,Gyr and 
an average [Fe/H] of $-1.7^{+0.07}_{-0.27}$ for the stellar population.

\item We analyze the structural properties of Seg\,3, using Plummer and 
exponential profiles.  We find structural parameters similar to those of 
\citet{belokurov10a}, except for the half-light radius which we find to be 
smaller by $\sim 30\%$.  Using a Salpeter IMF and CMD properties from 
our photometry, we find Seg\,3 has an absolute magnitude of 
$M_{V}=0.0\pm0.8$, making it the faintest stellar system known to date.

\item Examining 20 spectroscopic member stars within three half--light radii, we find a kinematically 
bound group of 19 member stars and identify one star which is an outlier of the 
velocity distribution (at 95\% confidence).  For this group of bound stars we infer a 
velocity dispersion of $1.2\pm2.6$ \kms, consistent with very low values for the 
true velocity dispersion.

\item Segue 3 is likely a faint stellar cluster with no significant dark matter 
content.  This is evidenced by our measurement of the velocity dispersion within 
three half--light radii, the (relatively) high average and low scatter in [Fe/H] 
values implied by isochrones, and small physical size of Seg\,3 relative 
to similar luminosity ultra--faint dwarf galaxies.

\item We find signs of possible mass loss in Segue 3, indicated by 
eleven candidate member stars outside of $3r_{1/2}$.
While these stars may originate from the Milky Way or nearby substructure, 
we estimate a Jacobi radius of $r_{\rm J}<5r_{1/2}$ and a half--mass relaxation time of 
$t_{\rm rh}\sim150$ Myr for Seg\,3, suggesting mass loss should be important.

\item Seg\,3 appears spatially coincident with an extended overdensity 
discovered by \citet{dejong10a}, which may be associated with the 
Hercules--Aquila cloud.  However, we find it is kinematically offset from the 
structure; evidenced by the low density of non-member stars within 
$\pm45$\kms\ from the systemic velocity of Seg\,3.  In our spectroscopic 
sample we identify a collection of stars with line--of--sight velocities 
$<200$\kms, which may be associated with the structure.  We find 
at most one star from this population may be misidentified as a member.

\end{itemize}
\phantom{need space between conclusions and acknowledgments}
\acknowledgments

RF and BW acknowledge support from NSF AST--0908193.  MG 
acknowledges support from NSF grant AST--0908752 and the Alfred 
P.~Sloan Foundation.  R. R. M. acknowledges support from the 
GEMINI-CONICYT Fund, allocated to the project N$^\circ$32080010. 
GDaC acknowledges research support in part through Australian 
Research Council Discovery Projects Grant DP0878137.  This paper 
includes data gathered with the 6.5 meter Magellan Telescopes located 
at Las Campanas Observatory, Chile.  Australian access to the Magellan 
Telescopes was supported through the National Collaborative Research 
Infrastructure Strategy of the Australian Federal Government.  Some of 
the data presented herein were obtained at the W.M. Keck Observatory, 
which is operated as a scientific partnership among the California Institute of
Technology, the University of California and the National Aeronautics
and Space Administration. The Observatory was made possible by the
generous financial support of the W.M. Keck Foundation. This research
has also made use of NASA's Astrophysics Data System Bibliographic
Services.

\bibliographystyle{apj}

\begin{thebibliography}{36}
\expandafter\ifx\csname natexlab\endcsname\relax\def\natexlab#1{#1}\fi

\bibitem[{{Belokurov} {et~al.}(2007){Belokurov}, {Evans}, {Bell}, {Irwin},
  {Hewett}, {Koposov}, {Rockosi}, {Gilmore}, {Zucker}, {Fellhauer},
  {Wilkinson}, {Bramich}, {Vidrih}, {Rix}, {Beers}, {Schneider}, {Barentine},
  {Brewington}, {Brinkmann}, {Harvanek}, {Krzesinski}, {Long}, {Pan},
  {Snedden}, {Malanushenko}, \& {Malanushenko}}]{belokurov07b}
{Belokurov}, V. {et~al.} 2007, \apjl, 657, L89

\bibitem[{{Belokurov} {et~al.}(2010){Belokurov}, {Walker}, {Evans}, {Gilmore},
  {Irwin}, {Just}, {Koposov}, {Mateo}, {Olszewski}, {Watkins}, \&
  {Wyrzykowski}}]{belokurov10a}
---. 2010, \apjl, 712, L103

\bibitem[{{Bertin}(2006)}]{bertin06a}
{Bertin}, E. 2006, in Astronomical Society of the Pacific Conference Series,
  Vol. 351, Astronomical Data Analysis Software and Systems XV, ed.
  {C.~Gabriel, C.~Arviset, D.~Ponz, \& S.~Enrique}, 112--+

\bibitem[{{Bertin} \& {Arnouts}(1996)}]{bertin96a}
{Bertin}, E., \& {Arnouts}, S. 1996, \aaps, 117, 393

\bibitem[{{Bertin} {et~al.}(2002){Bertin}, {Mellier}, {Radovich}, {Missonnier},
  {Didelon}, \& {Morin}}]{bertin02a}
{Bertin}, E., {Mellier}, Y., {Radovich}, M., {Missonnier}, G., {Didelon}, P.,
  \& {Morin}, B. 2002, in Astronomical Society of the Pacific Conference
  Series, Vol. 281, Astronomical Data Analysis Software and Systems XI, ed.
  {D.~A.~Bohlender, D.~Durand, \& T.~H.~Handley}, 228--+

\bibitem[{{Bullock} {et~al.}(2010){Bullock}, {Stewart}, {Kaplinghat},
  {Tollerud}, \& {Wolf}}]{bullock10}
{Bullock}, J.~S., {Stewart}, K.~R., {Kaplinghat}, M., {Tollerud}, E.~J., \&
  {Wolf}, J. 2010, \apj, 717, 1043

\bibitem[{{Carraro}(2009)}]{carraro09a}
{Carraro}, G. 2009, \aj, 137, 3809

\bibitem[{{Carraro} {et~al.}(2007){Carraro}, {Zinn}, \& {Moni
  Bidin}}]{carraro07a}
{Carraro}, G., {Zinn}, R., \& {Moni Bidin}, C. 2007, \aap, 466, 181

\bibitem[{{de Jong} {et~al.}(2010){de Jong}, {Yanny}, {Rix}, {Dolphin},
  {Martin}, \& {Beers}}]{dejong10a}
{de Jong}, J.~T.~A., {Yanny}, B., {Rix}, H., {Dolphin}, A.~E., {Martin}, N.~F.,
  \& {Beers}, T.~C. 2010, \apj, 714, 663

\bibitem[{{Dotter} {et~al.}(2008){Dotter}, {Chaboyer}, {Jevremovi{\'c}},
  {Kostov}, {Baron}, \& {Ferguson}}]{dotter08}
{Dotter}, A., {Chaboyer}, B., {Jevremovi{\'c}}, D., {Kostov}, V., {Baron}, E.,
  \& {Ferguson}, J.~W. 2008, \apjs, 178, 89

\bibitem[{{Faber} {et~al.}(2003){Faber}, {Phillips}, {Kibrick}, {Alcott},
  {Allen}, {Burrous}, {Cantrall}, {Clarke}, {Coil}, {Cowley}, {Davis}, {Deich},
  {Dietsch}, {Gilmore}, {Harper}, {Hilyard}, {Lewis}, {McVeigh}, {Newman},
  {Osborne}, {Schiavon}, {Stover}, {Tucker}, {Wallace}, {Wei}, {Wirth}, \&
  {Wright}}]{faber03a}
{Faber}, S.~M. {et~al.} 2003, in Presented at the Society of Photo-Optical
  Instrumentation Engineers (SPIE) Conference, Vol. 4841, Instrument Design and
  Performance for Optical/Infrared Ground-based Telescopes. Edited by Iye,
  Masanori; Moorwood, Alan F. M. Proceedings of the SPIE, Volume 4841, pp.
  1657-1669 (2003)., ed. M.~{Iye} \& A.~F.~M. {Moorwood}, 1657--1669

\bibitem[{{Forbes} \& {Bridges}(2010)}]{forbes10a}
{Forbes}, D.~A., \& {Bridges}, T. 2010, \mnras, 404, 1203

\bibitem[{{Frayn} \& {Gilmore}(2002)}]{frayn02a}
{Frayn}, C.~M., \& {Gilmore}, G.~F. 2002, \mnras, 337, 445

\bibitem[{{Frebel} {et~al.}(2010){Frebel}, {Simon}, {Geha}, \&
  {Willman}}]{frebel10a}
{Frebel}, A., {Simon}, J.~D., {Geha}, M., \& {Willman}, B. 2010, \apj, 708, 560

\bibitem[{{Frinchaboy} {et~al.}(2004){Frinchaboy}, {Majewski}, {Crane}, {Reid},
  {Rocha-Pinto}, {Phelps}, {Patterson}, \& {Mu{\~ n}oz}}]{frinchaboy04}
{Frinchaboy}, P.~M., {Majewski}, S.~R., {Crane}, J.~D., {Reid}, I.~N.,
  {Rocha-Pinto}, H.~J., {Phelps}, R.~L., {Patterson}, R.~J., \& {Mu{\~ n}oz},
  R.~R. 2004, \apjl, 602, L21

\bibitem[{{Gnedin} {et~al.}(1999){Gnedin}, {Lee}, \& {Ostriker}}]{gnedin99}
{Gnedin}, O.~Y., {Lee}, H.~M., \& {Ostriker}, J.~P. 1999, \apj, 522, 935

\bibitem[{{Gnedin} \& {Ostriker}(1997)}]{gnedin97}
{Gnedin}, O.~Y., \& {Ostriker}, J.~P. 1997, \apj, 474, 223

\bibitem[{{Harris}(1996)}]{harrisGCcat}
{Harris}, W.~E. 1996, \aj, 112, 1487

\bibitem[{{Kirby} {et~al.}(2011){Kirby}, {Lanfranchi}, {Simon}, {Cohen}, \&
  {Guhathakurta}}]{kirby11a}
{Kirby}, E.~N., {Lanfranchi}, G.~A., {Simon}, J.~D., {Cohen}, J.~G., \&
  {Guhathakurta}, P. 2011, \apj, 727, 78

\bibitem[{{Kirby} {et~al.}(2008){Kirby}, {Simon}, {Geha}, {Guhathakurta}, \&
  {Frebel}}]{kirby08a}
{Kirby}, E.~N., {Simon}, J.~D., {Geha}, M., {Guhathakurta}, P., \& {Frebel}, A.
  2008, \apjl, 685, L43

\bibitem[{{Koposov} {et~al.}(2007){Koposov}, {de Jong}, {Belokurov}, {Rix},
  {Zucker}, {Evans}, {Gilmore}, {Irwin}, \& {Bell}}]{koposov07b}
{Koposov}, S. {et~al.} 2007, \apj, 669, 337

\bibitem[{{Marigo} {et~al.}(2008){Marigo}, {Girardi}, {Bressan}, {Groenewegen},
  {Silva}, \& {Granato}}]{marigo08}
{Marigo}, P., {Girardi}, L., {Bressan}, A., {Groenewegen}, M.~A.~T., {Silva},
  L., \& {Granato}, G.~L. 2008, \aap, 482, 883

\bibitem[{{Martin} {et~al.}(2008){Martin}, {de Jong}, \& {Rix}}]{martin08b}
{Martin}, N.~F., {de Jong}, J.~T.~A., \& {Rix}, H.-W. 2008, \apj, 684, 1075

\bibitem[{{Mu{\~n}oz} {et~al.}(2010){Mu{\~n}oz}, {Geha}, \&
  {Willman}}]{munoz10a}
{Mu{\~n}oz}, R.~R., {Geha}, M., \& {Willman}, B. 2010, \aj, 140, 138

\bibitem[{{Niederste-Ostholt} {et~al.}(2010){Niederste-Ostholt}, {Belokurov},
  {Evans}, {Koposov}, {Gieles}, \& {Irwin}}]{NO10a}
{Niederste-Ostholt}, M., {Belokurov}, V., {Evans}, N.~W., {Koposov}, S.,
  {Gieles}, M., \& {Irwin}, M.~J. 2010, \mnras, 408, L66

\bibitem[{{Plummer}(1911)}]{Plummer11}
{Plummer}, H.~C. 1911, \mnras, 71, 460

\bibitem[{{Robin} {et~al.}(2003){Robin}, {Reyl{\'e}}, {Derri{\`e}re}, \&
  {Picaud}}]{robin03}
{Robin}, A.~C., {Reyl{\'e}}, C., {Derri{\`e}re}, S., \& {Picaud}, S. 2003,
  \aap, 409, 523

\bibitem[{{Simon} \& {Geha}(2007)}]{simon07a}
{Simon}, J.~D., \& {Geha}, M. 2007, \apj, 670, 313

\bibitem[{{Simon} {et~al.}(2011){Simon}, {Geha}, {Minor}, {Martinez}, {Kirby},
  {Bullock}, {Kaplinghat}, {Strigari}, {Willman}, {Choi}, {Tollerud}, \&
  {Wolf}}]{simon11a}
{Simon}, J.~D. {et~al.} 2011, \apj, 733, 46

\bibitem[{{Sohn} {et~al.}(2007){Sohn}, {Majewski}, {Mu{\~n}oz}, {Kunkel},
  {Johnston}, {Ostheimer}, {Guhathakurta}, {Patterson}, {Siegel}, \&
  {Cooper}}]{sohn07}
{Sohn}, S.~T. {et~al.} 2007, \apj, 663, 960

\bibitem[{{Spitzer} \& {Hart}(1971)}]{spitzer71}
{Spitzer}, Jr., L., \& {Hart}, M.~H. 1971, \apj, 164, 399

\bibitem[{{Stetson}(1987)}]{stetson87}
{Stetson}, P.~B. 1987, \pasp, 99, 191

\bibitem[{{Walker} {et~al.}(2006){Walker}, {Mateo}, {Olszewski}, {Bernstein},
  {Wang}, \& {Woodroofe}}]{walker06a}
{Walker}, M.~G., {Mateo}, M., {Olszewski}, E.~W., {Bernstein}, R., {Wang}, X.,
  \& {Woodroofe}, M. 2006, \aj, 131, 2114

\bibitem[{{Walsh} {et~al.}(2008){Walsh}, {Willman}, {Sand}, {Harris}, {Seth},
  {Zaritsky}, \& {Jerjen}}]{walsh08a}
{Walsh}, S.~M., {Willman}, B., {Sand}, D., {Harris}, J., {Seth}, A.,
  {Zaritsky}, D., \& {Jerjen}, H. 2008, \apj, 688, 245

\bibitem[{{Willman} {et~al.}(2011){Willman}, {Geha}, {Strader}, {Strigari},
  {Simon}, {Kirby}, \& {Warres}}]{willman11a}
{Willman}, B., {Geha}, M., {Strader}, J., {Strigari}, L.~E., {Simon}, J.~D.,
  {Kirby}, E., \& {Warres}, A. 2011, AJ submitted, arXiv:1007.3499

\bibitem[{{Wolf} {et~al.}(2010){Wolf}, {Martinez}, {Bullock}, {Kaplinghat},
  {Geha}, {Mu{\~n}oz}, {Simon}, \& {Avedo}}]{wolf10a}
{Wolf}, J., {Martinez}, G.~D., {Bullock}, J.~S., {Kaplinghat}, M., {Geha}, M.,
  {Mu{\~n}oz}, R.~R., {Simon}, J.~D., \& {Avedo}, F.~F. 2010, \mnras, 778

\end{thebibliography}

\clearpage

\begin{deluxetable}{lcc}
\tablewidth{0pt}
\tablecaption{Structural Parameters for Segue 3}
\tablehead{
 \multicolumn{3}{c}{\phantom{..}}\\
 \multicolumn{3}{c}{Exponential Profile}
}
\startdata
Right Ascention & $\alpha_{0}$ &\hspace{3.5pt}21:21:31.05\,$\pm\,1''.6$ \\
Declination & $\delta_{0}$ &+19:07:02.6\,$\pm\,2''.4$ \\
Half--light radius [$''$] & $r_{1/2}$ &\hspace{30pt} $28\pm8$ \\
Half--light radius [pc]\tablenotemark{a} & $ $ &\hspace{35.5pt} $2.2\pm0.7$ \\
Ellipticity &  $\epsilon$&\hspace{35pt} $0.24\pm0.14$ \\
Position Angle & $\theta$ &\hspace{35pt} $25^\circ \pm17^\circ$ \\
Number of stars\tablenotemark{b} & $N_{*}$ & \hspace{33pt}$65\pm6$ \\
Absolute Magnitude\tablenotemark{a}  & $M_{V}$& \hspace{29.5pt}$-0.06\pm0.78$ \\
Central surface brightness & $\mu_{V,0}$& \hspace{21pt}$23.9^{+1.0}_{-0.8}$  mag arcmin$^{-1}$ \\
\hline 
 \multicolumn{3}{c}{\phantom{--}}\\
 \multicolumn{3}{c}{Plummer Profile\vspace{2pt}}\\
\hline
Right Ascention & $\alpha_{0}$ &\hspace{3.5pt}21:21:31.02\,$\pm\,1''.8$ \\
Declination & $\delta_{0}$ &+19:07:03.7\,$\pm\,2''.6$ \\
Half--light radius [$''$]& $r_{1/2}$ & \hspace{34.pt}$26\pm5$ \\
Half--light radius [pc]\tablenotemark{a} & $ $ &\hspace{35.5pt} $2.1\pm0.4$ \\
Ellipticity &  $\epsilon$& \hspace{39pt}$0.23\pm0.11$ \\
Position Angle & $\theta$ & \hspace{39pt}$33^\circ \pm36^\circ$ \\
Number of stars\tablenotemark{b} & $N_{*}$ & \hspace{32pt}$64\pm6$ \\
Absolute Magnitude\tablenotemark{a} & $M_{V}$ & \hspace{28pt}$-0.04\pm0.78$ \\
Central surface brightness & $\mu_{V,0}$& \hspace{21pt}$24.1^{+1.0}_{-0.8}$  mag arcmin$^{-1}$
\enddata 
\tablenotetext{a}{Using a distance of 17 kpc for Segue~3.}
\tablenotetext{b}{For stars with $r < 22.5$ mag.}
\label{tab:struct}
\end{deluxetable}

\begin{deluxetable}{ccccccccc}
\tabletypesize{\scriptsize}
\tablecaption{Photometric and Kinematic Data for Keck/DEIMOS sample -- \newline
 I. Candidate Members.}
\tablewidth{0pt}
\tablehead{
\colhead{} &
\colhead{} &
\colhead{} &
\colhead{} &
\colhead{Radial} &
\colhead{} &
\colhead{} &
\colhead{} & \\
\colhead{ID} &
\colhead{Date} &
\colhead{$\alpha$ (J2000)} &
\colhead{$\delta$ (J2000)} &
\colhead{Distance} &
\colhead{$r$} &
\colhead{$(g-r)$} &
\colhead{$v_{\rm helio}$} & \\
\colhead{}&
\colhead{y.m.d}&
\colhead{(h$\,$ $\,$ m$\,$ $\,$s)} &
\colhead{($^\circ\,$ $\,'\,$ $\,''$)} &
\colhead{($'$)} &
\colhead{(mag)} &
\colhead{(mag)} &
\colhead{(\kms)} &
 }
\startdata
1 & 2009.11.16 & 21:21:31.4 & +19:07:00.3 & 0.11 &  20.1 &  0.23 & $-162.0\pm4.5$ \\
 2 & combined & 21:21:31.5 & +19:07:04.0 & 0.11 &  20.2 &  0.22 & $-158.9\pm3.2$ \\
   & 2009.11.16 &   &   & &   & & $\hspace{4.3pt}-179.3\pm10.6$ \\
   & 2010.05.16 &   &   & &   & & $-157.9\pm3.2$ \\
 3 & 2009.11.16 & 21:21:31.5 & +19:07:09.6 & 0.17 &  20.7 &  0.25 & $-160.9\pm7.6$ \\
 4 & 2009.11.16 & 21:21:30.3 & +19:06:58.5 & 0.18 &  20.0 &  0.22 & $-170.0\pm3.8$ \\
 5 & combined & 21:21:30.4 & +19:07:13.4 & 0.23 &  21.2 &  0.30 & $-168.9\pm5.4$ \\
     & 2009.11.16 &  & & &  &  & $-168.7\pm6.8$ \\
     & 2010.05.16 &  & & &  &  & $-169.2\pm8.0$ \\
 6 & 2009.11.16 & 21:21:29.9 & +19:07:07.7 & 0.28 &  20.6 &  0.23 & $-153.3\pm5.9$ \\
 7 & 2010.05.16 & 21:21:30.0 & +19:06:52.2 & 0.28 &  21.9 &  0.41 & $-171.3\pm9.6$ \\
 8 & combined & 21:21:30.3 & +19:06:45.6 & 0.32 &  20.4 &  0.22 & $-167.6\pm4.7$ \\
    & 2009.11.16 &  &  &  &  &  & $-170.1\pm9.1$ \\
    & 2010.05.16 &  &  &  &  &  & $-166.1\pm5.2$ \\
 9 & combined & 21:21:32.7 & +19:06:57.4 & 0.42 &  20.5 &  0.27 & $-161.2\pm3.1$ \\
    & 2009.11.16 & &  & & & & $-176.3\pm3.8$ \\
    & 2010.05.16 & &  & & & & $-146.3\pm3.8$ \\
10 & 2009.11.16 & 21:21:29.4 & +19:07:12.5 & 0.42 &  19.8 &  0.20 & $-163.2\pm3.0$ \\
11 & 2010.05.16 & 21:21:29.0 & +19:07:19.6 & 0.56 &  22.1 &  0.49 & $-172.5\pm7.8$ \\
12 & 2009.11.16 & 21:21:29.8 & +19:07:30.4 & 0.56 &  21.1 &  0.28 & $-175.7\pm6.3$ \\
13 & combined & 21:21:33.1 & +19:06:40.3 & 0.61 &  20.5 &  0.23 & $-166.1\pm4.2$ \\
    & 2009.11.16 &  &  &  &   &  & $\hspace{4.3pt}-155.7\pm11.3$ \\
    & 2010.05.16 &  &  &  &   &  & $-167.4\pm4.4$ \\
14 & 2009.11.16 & 21:21:30.8 & +19:06:21.5 & 0.68 &  20.4 &  0.23 & $-155.2\pm4.0$ \\
15 & 2009.11.16 & 21:21:31.1 & +19:07:51.9 & 0.83 &  19.4 &  0.28 & $-163.5\pm2.8$ \\
16 & 2009.11.16 & 21:21:32.7 & +19:07:47.9 & 0.86 &  21.6 &  0.38 & $-169.9\pm5.3$ \\
17 & combined & 21:21:33.6 & +19:06:23.8 & 0.88 &  20.0 &  0.22 & $-162.7\pm2.7$ \\
     & 2009.11.16 & &  & &  &  & $-158.6\pm3.3$ \\
     & 2010.05.16 & &  & &  &  & $-165.4\pm3.0$ \\
18 & 2010.05.16 & 21:21:29.1 & +19:06:15.3 & 0.89 &  21.2 &  0.33 & $-167.7\pm4.6$ \\
19 & 2009.11.16 & 21:21:27.1 & +19:07:12.5 & 0.94 &  21.4 &  0.32 & $-165.3\pm3.6$ \\
20 & combined & 21:21:26.4 & +19:06:35.6 & 1.18 &  20.4 &  0.25 & $-182.4\pm3.0$ \\
      & 2009.11.16 & & && & & $-184.6\pm4.0$ \\
      & 2010.05.16 & & && & & $-181.1\pm3.8$ \\
21 & combined & 21:21:31.0 & +19:08:16.6 & 1.24 &  20.5 &  0.24 & $-166.6\pm3.4$ \\
      & 2009.11.16 && &  &  & & $-167.7\pm4.5$ \\
      & 2010.05.16 && &  &  & & $-165.9\pm4.0$ \\
22 & combined & 21:21:38.1 & +19:07:33.5 & 1.75 &  21.7 &  0.36 & $-183.1\pm4.7$ \\
      & 2009.11.16 & & & & & & $\hspace{4.3pt}-167.9\pm10.8$ \\
      & 2010.05.16 & & & & & & $-186.0\pm5.1$ \\
23 & combined & 21:21:32.0 & +19:08:52.5 & 1.86 &  21.2 &  0.31 & $-167.6\pm4.5$ \\
      & 2009.11.16 &&  & &  && $\hspace{4.3pt}-158.5\pm11.4$ \\
      & 2010.05.16 &&  & &  && $-168.8\pm4.7$ \\
24 & combined & 21:21:36.2 & +19:08:27.2 & 1.88 &  21.6 &  0.36 & $-166.8\pm5.5$ \\
      & 2009.11.16 & & & & & & $-162.4\pm6.2$ \\
      & 2010.05.16 & & & & & & $\hspace{4.3pt}-180.7\pm10.6$ \\
25 & 2010.05.16 & 21:21:33.2 & +19:05:05.1 & 2.02 &  19.2 &  0.34 & $-169.8\pm2.4$ \\
26 & 2010.05.16 & 21:21:39.0 & +19:07:51.7 & 2.06 &  22.0 &  0.39 & $-152.9\pm7.3$ \\
27 & combined & 21:21:32.8 & +19:10:02.8 & 3.04 &  19.4 &  0.32 & $-183.3\pm2.3$ \\
     & 2009.11.16 &  & &&  &  & $-180.2\pm2.9$ \\
     & 2010.05.16 &  & &&  &  & $-184.0\pm2.4$ \\
28 & 2009.11.16 & 21:21:16.9 & +19:06:32.8 & 3.37 &  21.4 &  0.35 & $-159.5\pm6.7$ \\
29 & 2010.05.16 & 21:21:24.7 & +19:03:52.9 & 3.49 &  21.0 &  0.28 & $-175.6\pm7.9$ \\
30 & 2009.11.16 & 21:21:15.9 & +19:05:25.4 & 3.92 &  20.7 &  0.27 & $-173.6\pm4.8$ \\
31 & combined & 21:21:31.2 & +19:12:36.2 & 5.57 &  19.2 &  0.30 & $-153.5\pm2.4$ \\
     & 2009.11.16 & & &  &&& $-156.9\pm2.7$ \\
     & 2010.05.16 & & &  &&& $-151.8\pm2.5$ \\
32 & 2009.11.16 & 21:21:05.8 & +19:05:33.3 & 6.14 &  19.9 &  0.18 & $-169.6\pm4.0$ \\
\enddata
\label{tab:mem}
\tablecomments{Velocity error bars were determined from measurement overlaps
  as discussed in Section \ref{ssec:spect}.}

\end{deluxetable} 
\clearpage

\begin{longtable}{ccccccccc}
\tabletypesize{\scriptsize}
\tablecaption{Photometric and Kinematic Data for Keck/DEIMOS sample -- \newline
II. Non-members.}
\tablewidth{0pt}
\tablehead{
\colhead{} &
\colhead{} &
\colhead{} &
\colhead{} &
\colhead{Radial} &
\colhead{} &
\colhead{} &
\colhead{} & \\
\colhead{ID} &
\colhead{Date} &
\colhead{$\alpha$ (J2000)} &
\colhead{$\delta$ (J2000)} &
\colhead{Distance} &
\colhead{$r$} &
\colhead{$(g-r)$} &
\colhead{$v_{\rm helio}$} & \\
\colhead{}&
\colhead{y.m.d}&
\colhead{(h$\,$ $\,$ m$\,$ $\,$s)} &
\colhead{($^\circ\,$ $\,'\,$ $\,''$)} &
\colhead{($'$)} &
\colhead{(mag)} &
\colhead{(mag)} &
\colhead{(\kms)} &
 }
\startdata
 33 & 2009-11-16 & 21:21:33.9 & +19:07:21.5 & 0.75 &  16.7 &  0.51 & $ -60.6\pm2.2$ \\
 34 & 2009-11-16 & 21:21:27.9 & +19:07:19.3 & 0.78 &  21.5 &  0.47 & $-237.3\pm4.5\hspace{4.5pt}$ \\
 35 & 2009-11-16 & 21:21:32.7 & +19:06:12.8 & 0.91 &  17.8 &  0.44 & $ -54.9\pm2.2$ \\
 36 & 2009-11-16 & 21:21:34.6 & +19:06:33.1 & 0.98 &  16.6 &  0.53 & $ -96.6\pm2.2$ \\
 37 & combined & 21:21:32.0 & +19:08:02.5 & 1.04 &  19.8 &   0.25 & $-161.8\pm2.6\hspace{4.5pt}$ \\
      & 2009.11.16 &  &  & &  &   & $-159.6\pm3.1\hspace{4.5pt}$ \\
      & 2010.05.16 &  &  & &  &   & $-163.4\pm2.9\hspace{4.5pt}$ \\
 38 & combined & 21:21:28.4 & +19:06:00.7 & 1.20 &  19.3 &   0.48 & $-171.7\pm2.4\hspace{4.5pt}$ \\
      & 2009.11.16 & &  && & & $-171.0\pm2.6\hspace{4.5pt}$ \\
      & 2010.05.16 & &  && & & $-172.3\pm2.6\hspace{4.5pt}$ \\
 39 & 2009-11-16 & 21:21:32.3 & +19:05:52.2 & 1.20 &  19.2 &  0.49 & $-163.3\pm2.3\hspace{4.5pt}$ \\
 40 & 2009-11-16 & 21:21:27.6 & +19:06:06.1 & 1.23 &  21.5 &  0.49 & $-353.7\pm4.7\hspace{4.5pt}$ \\
 41 & 2009-11-16 & 21:21:34.0 & +19:05:56.9 & 1.29 &  16.1 &  0.72 & $ -35.1\pm2.2$ \\
 42 & 2009-11-16 & 21:21:29.6 & +19:05:25.3 & 1.64 &  18.6 &  0.51 & $ -79.9\pm2.2$ \\
 43 & 2009-11-16 & 21:21:25.3 & +19:06:05.9 & 1.65 &  16.9 &  0.44 & $ -22.0\pm2.2$ \\
 44 & 2009-11-16 & 21:21:36.0 & +19:05:44.2 & 1.75 &  16.6 &  0.43 & $\phantom{-}  28.8\pm2.2$ \\
 45 & 2009-11-16 & 21:21:38.8 & +19:07:10.0 & 1.84 &  21.2 &  0.41 & $-166.1\pm6.2\hspace{4.5pt}$ \\
 46 & 2009-11-16 & 21:21:39.6 & +19:06:14.2 & 2.19 &  18.5 &  0.54 & $ -49.0\pm2.2$ \\
 47 & 2009-11-16 & 21:21:21.4 & +19:07:26.6 & 2.31 &  16.1 &  0.56 & $\hspace{4.0pt} -3.6\pm2.2$ \\
 48 & 2009-11-16 & 21:21:22.5 & +19:08:12.6 & 2.33 &  16.3 &  0.54 & $ -14.5\pm2.2$ \\
 49 & 2010-05-16 & 21:21:29.2 & +19:09:26.9 & 2.45 &  19.4 &  0.48 & $ -98.2\pm2.3$ \\
 50 & 2009-11-16 & 21:21:42.6 & +19:07:08.1 & 2.74 &  19.0 &  0.34 & $-120.5\pm2.4\hspace{4.5pt}$ \\
 51 & 2009-11-16 & 21:21:38.3 & +19:04:46.0 & 2.84 &  17.8 &  0.39 & $\phantom{-}\hspace{4.0pt}  3.2\pm2.2$ \\
 52 & 2010-05-16 & 21:21:25.9 & +19:09:42.5 & 2.94 &  20.9 &  0.44 & $\phantom{-}  27.7\pm5.2$ \\
 53 & 2010-05-16 & 21:21:32.7 & +19:04:02.1 & 3.03 &  17.4 &  0.36 & $-109.6\pm2.2\hspace{4.5pt}$ \\
 54 & 2010-05-16 & 21:21:20.7 & +19:05:11.9 & 3.05 &  18.5 &  0.49 & $\hspace{4.0pt} -8.4\pm2.2$ \\
 55 & 2009-11-16 & 21:21:44.2 & +19:07:35.6 & 3.16 &  19.4 &  0.37 & $-221.1\pm2.5\hspace{4.5pt}$ \\
 56 & 2009-11-16 & 21:21:28.7 & +19:10:13.1 & 3.23 &  18.3 &  0.42 & $ -40.2\pm2.3$ \\
 57 & 2009-11-16 & 21:21:40.4 & +19:04:28.4 & 3.39 &  18.4 &  0.53 & $ -42.7\pm2.2$ \\
 58 & 2009-11-16 & 21:21:16.7 & +19:06:33.2 & 3.40 &  21.3 &  1.17 & $\phantom{-}  25.7\pm3.2$ \\
 59 & 2009-11-16 & 21:21:24.4 & +19:03:58.8 & 3.42 &  18.7 &  0.51 & $ -31.3\pm2.3$ \\
 60 & 2009-11-16 & 21:21:43.3 & +19:08:56.1 & 3.47 &  17.8 &  0.37 & $\phantom{-}\hspace{4.0pt}  2.7\pm2.2$ \\
 61 & 2010-05-16 & 21:21:32.9 & +19:10:29.3 & 3.48 &  19.8 &  1.15 & $ -19.5\pm7.1$ \\
 62 & 2009-11-16 & 21:21:39.3 & +19:04:06.2 & 3.53 &  19.0 &  0.37 & $-323.2\pm2.4\hspace{4.5pt}$ \\
 63 & 2009-11-16 & 21:21:33.9 & +19:10:30.8 & 3.55 &  17.1 &  0.43 & $ -45.6\pm2.2$ \\
 64 & 2009-11-16 & 21:21:34.8 & +19:03:30.9 & 3.63 &  18.3 &  0.37 & $-115.2\pm2.3\hspace{4.5pt}$ \\
 65 & 2009-11-16 & 21:21:30.6 & +19:10:43.4 & 3.69 &  19.5 &  0.46 & $ -67.3\pm2.3$ \\
 66 & 2009-11-16 & 21:21:15.0 & +19:06:31.5 & 3.82 &  16.4 &  0.48 & $\phantom{-}\hspace{4.0pt}  6.6\pm2.2$ \\
 67 & 2009-11-16 & 21:21:29.9 & +19:10:54.1 & 3.88 &  19.8 &  0.33 & $-300.6\pm3.3\hspace{4.5pt}$ \\
 68 & 2009-11-16 & 21:21:14.2 & +19:07:52.7 & 4.07 &  17.1 &  0.51 & $ -89.6\pm2.2$ \\
 69 & 2009-11-16 & 21:21:47.5 & +19:08:21.6 & 4.11 &  17.4 &  0.58 & $ -69.4\pm2.3$ \\
 70 & 2009-11-16 & 21:21:22.7 & +19:03:25.2 & 4.11 &  18.3 &  0.38 & $-113.2\pm2.5\hspace{4.5pt}$ \\
 71 & 2009-11-16 & 21:21:38.9 & +19:10:49.6 & 4.22 &  18.0 &  0.55 & $\phantom{-}\hspace{4.0pt}  2.4\pm2.2$ \\
 72 & 2009-11-16 & 21:21:22.7 & +19:03:15.4 & 4.25 &  20.7 &  0.81 & $-117.3\pm2.8\hspace{4.5pt}$ \\
 73 & combined & 21:21:37.6 & +19:03:03.9 & 4.26 &  17.1 &   0.37 & $-160.6\pm2.2\hspace{4.5pt}$ \\
      & 2009.11.16 & & &  &  & & $-160.7\pm2.2\hspace{4.5pt}$ \\
      & 2010.05.16 & & &  &  & & $-160.6\pm2.2\hspace{4.5pt}$ \\
 74 & 2010-05-16 & 21:21:35.1 & +19:11:11.6 & 4.27 &  22.2 &  0.58 & $-331.8\pm5.1\hspace{4.5pt}$ \\
 75 & 2010-05-16 & 21:21:34.4 & +19:02:48.1 & 4.31 &  18.7 &  0.34 & $-274.2\pm2.3\hspace{4.5pt}$ \\
 76 & 2009-11-16 & 21:21:48.0 & +19:05:20.7 & 4.36 &  19.5 &  0.39 & $ -97.4\pm2.4$ \\
 77 & 2009-11-16 & 21:21:26.9 & +19:02:44.6 & 4.40 &  16.4 &  0.63 & $ -19.4\pm2.2$ \\
 78 & 2010-05-16 & 21:21:23.4 & +19:11:06.5 & 4.45 &  16.2 &  0.51 & $ -30.9\pm2.2$ \\
 79 & 2009-11-16 & 21:21:34.1 & +19:02:37.2 & 4.47 &  18.2 &  0.48 & $ -83.2\pm2.3$ \\
 80 & 2009-11-16 & 21:21:13.1 & +19:05:26.9 & 4.52 &  18.8 &  0.52 & $ -39.1\pm2.3$ \\
 81 & 2009-11-16 & 21:21:35.5 & +19:11:26.1 & 4.53 &  16.9 &  0.59 & $ -62.8\pm2.2$ \\
 82 & 2009-11-16 & 21:21:11.8 & +19:07:03.2 & 4.55 &  18.3 &  0.54 & $ -90.0\pm2.2$ \\
 83 & 2009-11-16 & 21:21:49.8 & +19:08:15.6 & 4.61 &  18.3 &  0.45 & $ -76.6\pm2.2$ \\
 84 & 2009-11-16 & 21:21:12.5 & +19:05:25.2 & 4.66 &  17.9 &  0.47 & $ -71.9\pm2.2$ \\
 85 & 2009-11-16 & 21:21:11.1 & +19:06:23.1 & 4.75 &  21.4 &  0.41 & $-162.4\pm3.9\hspace{4.5pt}$ \\
 86 & 2010-05-16 & 21:21:38.7 & +19:11:29.3 & 4.81 &  20.0 &  0.30 & $-177.3\pm2.9\hspace{4.5pt}$ \\
 87 & 2009-11-16 & 21:21:29.9 & +19:11:52.0 & 4.84 &  19.8 &  0.28 & $-239.4\pm2.8\hspace{4.5pt}$ \\
 88 & 2009-11-16 & 21:21:39.2 & +19:02:32.3 & 4.89 &  17.5 &  0.32 & $ -75.9\pm2.2$ \\
 89 & 2010-05-16 & 21:21:22.7 & +19:11:37.5 & 4.99 &  17.1 &  0.55 & $\phantom{-}  29.8\pm2.2$ \\
 90 & 2009-11-16 & 21:21:31.1 & +19:12:05.1 & 5.05 &  17.0 &  0.48 & $ -39.3\pm2.2$ \\
 91 & 2010-05-16 & 21:21:22.5 & +19:11:42.4 & 5.09 &  18.6 &  0.63 & $ -23.7\pm2.2$ \\
 92 & 2010-05-16 & 21:21:22.5 & +19:11:45.0 & 5.13 &  16.7 &  0.63 & $ -54.1\pm2.2$ \\
 93 & 2009-11-16 & 21:21:40.9 & +19:11:41.9 & 5.21 &  17.3 &  0.59 & $ -30.0\pm2.2$ \\
 94 & 2009-11-16 & 21:21:52.9 & +19:07:43.9 & 5.23 &  15.9 &  0.67 & $ \phantom{-} 30.2\pm2.2$ \\
 95 & 2009-11-16 & 21:21:08.8 & +19:07:14.0 & 5.24 &  21.4 &  0.44 & $-172.2\pm9.6\hspace{4.5pt}$ \\
 96 & 2009-11-16 & 21:21:52.6 & +19:05:42.5 & 5.27 &  20.9 &  0.31 & $-298.4\pm6.9\hspace{4.5pt}$ \\
 97 & 2010-05-16 & 21:21:22.3 & +19:02:02.0 & 5.40 &  18.2 &  0.45 & $\hspace{4.0pt} -4.6\pm2.2$ \\
 98 & 2010-05-16 & 21:21:22.5 & +19:12:03.4 & 5.41 &  21.6 &  1.01 & $ -48.1\pm5.6$ \\
 99 & 2009-11-16 & 21:21:32.5 & +19:12:26.7 & 5.42 &  18.0 &  0.32 & $ -27.5\pm2.2$ \\
100 & 2009-11-16 & 21:21:08.0 & +19:06:54.5 & 5.44 &  18.6 &  0.45 & $ -37.8\pm2.3$\\
101 & 2009-11-16 & 21:21:10.5 & +19:09:40.1 & 5.51 &  20.7 &  0.35 & $-133.1\pm3.7\hspace{4.5pt}$ \\
102 & 2010-05-16 & 21:21:37.1 & +19:01:34.1 & 5.65 &  20.9 &  0.31 & $-438.2\pm4.4\hspace{4.5pt}$ \\
103 & 2009-11-16 & 21:21:07.5 & +19:08:45.9 & 5.81 &  17.5 &  0.55 & $ -31.7\pm2.3$ \\
104 & 2010-05-16 & 21:21:24.9 & +19:01:22.3 & 5.84 &  17.6 &  0.48 & $ -55.9\pm2.2$ \\
105 & 2009-11-16 & 21:21:56.9 & +19:05:50.4 & 6.23 &  16.3 &  0.53 & $ -17.7\pm2.2$ \\
106 & combined & 21:21:28.7 & +19:13:19.1 & 6.31 &  21.2 &   0.17 & $-284.5\pm5.0\hspace{4.5pt}$ \\
        & 2009.11.16 &&  &  & &  & $\hspace{4.3pt}-297.5\pm11.8\hspace{4.5pt}$ \\
        & 2010.05.16 &&  &  & &  & $-282.2\pm5.3\hspace{4.5pt}$ \\
107 & 2009-11-16 & 21:21:57.5 & +19:08:07.8 & 6.36 &  17.9 &  0.57 & $ -55.7\pm2.2$ \\
108 & 2010-05-16 & 21:21:33.0 & +19:13:33.1 & 6.54 &  16.6 &  0.51 & $ -21.3\pm2.2$ \\
109 & 2009-11-16 & 21:21:56.5 & +19:09:40.7 & 6.58 &  18.5 &  0.55 & $ -55.7\pm2.2$ \\
110 & 2010-05-16 & 21:21:28.3 & +19:13:48.3 & 6.80 &  17.3 &  0.38 & $\phantom{-}\hspace{4.0pt}  0.5\pm2.2$ \\
111 & 2010-05-16 & 21:21:39.5 & +19:00:31.0 & 6.82 &  16.6 &  0.55 & $ -27.9\pm2.7$ \\
112 & 2009-11-16 & 21:21:59.9 & +19:07:31.5 & 6.85 &  17.1 &  0.44 & $\hspace{4.0pt} -2.2\pm2.2$ \\
113 & 2009-11-16 & 21:21:29.3 & +19:13:53.3 & 6.87 &  17.1 &  0.64 & $ -41.3\pm2.2$ \\
114 & 2009-11-16 & 21:21:24.8 & +19:13:44.8 & 6.87 &  16.7 &  0.41 & $ -97.1\pm2.2$ \\
115 & 2009-11-16 & 21:21:32.6 & +19:13:59.2 & 6.96 &  17.1 &  0.38 & $ -10.4\pm2.2$ \\
116 & 2009-11-16 & 21:21:58.4 & +19:09:41.7 & 7.01 &  17.0 &  0.60 & $\phantom{-}\hspace{4.0pt}  6.2\pm2.2$ \\
117 & 2009-11-16 & 21:22:01.1 & +19:06:38.3 & 7.11 &  17.8 &  0.50 & $\phantom{-}\hspace{4.0pt}  3.4\pm2.2$ \\
118 & 2010-05-16 & 21:21:30.6 & +19:14:09.9 & 7.13 &  16.1 &  0.48 & $ -47.2\pm2.2$ \\
119 & 2009-11-16 & 21:21:37.7 & +19:14:06.8 & 7.25 &  18.6 &  0.46 & $ -44.8\pm2.3$ \\
120 & 2009-11-16 & 21:21:39.0 & +19:14:22.4 & 7.58 &  17.8 &  0.49 & $ -34.9\pm2.2$ \\
121 & 2009-11-16 & 21:22:03.7 & +19:06:59.6 & 7.72 &  18.6 &  0.32 & $ -62.7\pm2.3$ \\
122 & 2009-11-16 & 21:22:03.3 & +19: 8:35.9 & 7.80 &  16.3 &  0.52 & $\phantom{-}  14.0\pm2.2$ \\
123 & 2010-05-16 & 21:21:21.1 & +19:14:48.2 & 8.11 &  18.2 &  0.44 & $ -26.4\pm2.5$ \\
124 & 2009-11-16 & 21:21:39.6 & +19:14:54.8 & 8.14 &  21.3 &  0.24 & $-236.6\pm5.9\hspace{4.5pt}$ \\
125 & 2010-05-16 & 21:21:30.8 & +19:15:12.7 & 8.18 &  18.5 &  0.48 & $\phantom{-}\hspace{4.0pt}  6.2\pm2.2$ \\
126 & 2009-11-16 & 21:21:39.5 & +19:15: 1.9 & 8.25 &  18.3 &  0.46 & $-273.7\pm2.3\hspace{4.5pt}$ \\
127 & 2009-11-16 & 21:21:24.0 & +19:15:26.8 & 8.57 &  18.2 &  0.63 & $ -17.8\pm2.2$ \\
128 & 2009-11-16 & 21:21:37.1 & +19:15:40.8 & 8.77 &  18.4 &  0.35 & $ -26.4\pm2.3$ \\
129 & 2009-11-16 & 21:22: 8.7 & +19: 7:42.4 & 8.94 &  17.9 &  0.41 & $ -51.5\pm2.2$ \\
130 & 2009-11-16 & 21:21:38.0 & +19:15:56.8 & 9.07 &  18.0 &  0.34 & $ -53.9\pm2.2$ \\
131 & 2010-05-16 & 21:21:30.0 & +19:16:21.1 & 9.32 &  17.6 &  0.54 & $ -32.6\pm2.3$ \\
132 & 2009-11-16 & 21:21:38.1 & +19:16:18.5 & 9.42 &  18.7 &  0.41 & $-285.0\pm2.4\hspace{4.5pt}$ 
\label{tab:nonmem}
\end{longtable}  

 \footnotesize \tablecomments{Velocity error bars were determined from measurement overlaps
  as discussed in Section \ref{ssec:spect}.}

\end{document}